\documentclass[prb,showpacs,twocolumn,amsmath,amssymb]{revtex4}
\usepackage{graphicx}
\usepackage{dcolumn}
\usepackage{bm}
\begin{document}
\par
\title{Response of thin-film SQUIDs to applied fields and vortex fields:
Linear SQUIDs}
\author{John R. Clem}
\affiliation{Ames Laboratory - DOE and Department of Physics and
Astronomy, Iowa State University, Ames Iowa 50011 }
\author{Ernst Helmut Brandt}
\affiliation{Max-Planck-Institut f\"ur Metallforschung,
      D-70506 Stuttgart, Germany}
\date{\today}
\begin{abstract} 
In this paper we analyze the properties of a dc SQUID  when the London 
penetration
depth $\lambda$ is larger than the superconducting film thickness $d$. 
We present equations that govern the static behavior for arbitrary
values of
$\Lambda = \lambda^2/d$ relative to the linear dimensions of the
SQUID.  The SQUID's critical current $I_c$ depends upon the
{\it effective flux} $\Phi$,  
the magnetic flux through a contour surrounding the central hole plus a
term proportional to the line integral of the current density around
this contour. While it is well known that the SQUID inductance  depends
upon 
$\Lambda$, we show here that the focusing of  magnetic flux from
applied fields and vortex-generated fields into the central hole of the
SQUID also depends upon 
$\Lambda$.
We apply this formalism to the simplest case of a
linear SQUID of width $2w$, consisting
of a coplanar pair of long superconducting strips of separation
$2a$, connected  by two
small Josephson junctions to a superconducting current-input lead at 
one end and by a
superconducting lead at the other end.  The central region of this SQUID
shares many properties with a superconducting coplanar stripline.
We calculate magnetic-field and current-density profiles, the
inductance (including both geometric and kinetic
inductances), magnetic moments, and the effective area as a function of
$\Lambda/w$ and $a/w$. 
\end{abstract}
\maketitle
\section{Introduction}

The research in this paper has been motivated by several important recent
developments in superconductivity:  (a) the fabrication of thin-film
SQUIDs (superconducting quantum interference devices) made of high-$T_c$
superconductors,
\cite{Koelle99} (b) the study of noise generated by vortices in active
and passive superconducting devices,
\cite{Ferrari91,Miklich94,Glyantsev96,Koelle99,Humphreys99,Straub01,
Woerdenweber02,Doenitz04,Nowak05,Kuriki04} 
and (c) line-width reduction in superconducting devices to eliminate
noise due to vortices trapped  during cooldown in the earth's magnetic
field.\cite{ClemUnpub,Dantsker96,Dantsker97,Mitchell03,Stan04,Du04} 

The fact that the London penetration depth $\lambda$ increases as $T$
increases and diverges at $T_c$ is an important consideration for
high-$T_c$ SQUIDs operated at liquid-nitrogen temperature.  When $\lambda$
is larger than the film thickness $d$, the physical length that enters
the equations governing the spatial variation of currents and fields is
the Pearl length\cite{Pearl64} $\Lambda = \lambda^2/d$.  Accordingly, the
equations governing the behavior of active or passive thin-film
superconducting devices depend upon the ratio of
$\Lambda$ to the linear dimensions of the device. In particular, the
equations governing SQUIDs involve not just the magnetic
flux up through a contour within the SQUID, but the effective
flux $\Phi$, which is the sum of the magnetic
flux and a term proportional
to the line integral of the current density around the same
contour.  While the effective flux $\Phi$ is similar to London's
fluxoid, \cite{London61} which is quantized in multiples of the
superconducting flux quantum $\phi_0 = h/2e$, we  show in the next
section that
$\Phi$ is not quantized.  We also give in Sec.\ II the basic equations,
valid for
any value of $\Lambda$, that
govern the behavior of a dc SQUID.

A vortex trapped in the body of the SQUID during cooldown
through the superconducting transition temperature $T_c$ in an ambient
magnetic field generates a magnetic field and a screening current that
together make a sizable vortex-position-dependent contribution 
$\Phi_{\rm v}$
to the effective flux
$\Phi$.  If such a vortex remains fixed in position and the temperature
remains constant, this simply produces a harmless bias in
$\Phi$.  On the other hand, both vortex motion due to
thermal agitation and temperature fluctuations generate
corresponding fluctuations in $\Phi_{\rm v}$ and noise in the SQUID output. 
In Sec.\ II we show that to calculate $\Phi_{\rm v}$, it is not necessary
to calculate the spatial dependence of the vortex-generated fields and
currents.  Instead, one may determine $\Phi_{\rm v}$ with the help of the
sheet-current distribution of a circulating current in the absence of the
vortex.               

In Sec.\ III we apply the basic equations of Sec.\ II to calculate the 
properties of a model linear SQUID, which has the basic topology of a
SQUID but is greatly stretched along one axis, such
that the central portion resembles a coplanar stripline.  The advantage
of using  such
a model is that simple analytical results can be derived that closely
approximate the exact numerically calculated quantities in the
appropriate limits.     In addition to
calculating the field and current distributions for several values of
$\Lambda$, we calculate the total inductance, geometric inductance,
kinetic inductance, and magnetic moment when the SQUID carries a
circulating current.   We calculate the field and current
distributions, magnetic moment, and effective area 
$A_{\rm eff} = \Phi_{\rm f}/B_{\rm a}$ when a perpendicular magnetic
induction
$B_{\rm a}$ is applied and 
the effective flux
$\Phi_{\rm f}$ is focused into the SQUID. 
Finally, we calculate the field and current distributions and the magnetic
moment for the zero-fluxoid state when the junctions are short-circuited
and the sample remains in the state with $\Phi = 0$ when a perpendicular
magnetic induction
$B_{\rm a}$ is applied.

In Sec.\ IV, we present a brief summary of our results.

\section{Basic equations}

Our purpose in this section is to derive general equations that
govern the behavior of a dc SQUID consisting of thin
superconducting films of thickness $d$ less than the weak-field
London penetration depth
$\lambda$, such that the fields and currents are governed by the
two-dimensional screening length or Pearl length~\cite{Pearl64}
$\Lambda=\lambda^2/d$.  We calculate both the current $I=I_1+I_2$
through the SQUID [see Fig.\ 1] and the circulating
current\cite{foot1}
$I_{\rm d}=(I_2-I_1)/2$ and describe how to calculate the critical
current $I_{\rm c}$ of the SQUID for arbitrary values of the SQUID's
inductance $L$. For this case, the contributions from line
integrals of the current density to the {\it effective flux} in the
hole cannot be neglected, and the kinetic inductance makes a
significant contribution to $L$. When  a perpendicular  magnetic
induction $B_{\rm a }$ is applied,  we calculate how much magnetic
flux is focused into the SQUID's hole; this flux also can be
expressed in terms of the  effective area\cite{Ketchen85} of
the hole.  We also show how to calculate how much magnetic flux
generated by a vortex in the main body of the SQUID is focused into
the hole.

Consider a dc SQUID in the
$xy$ plane, as sketched in Fig.\ 1.  We suppose that the SQUID is
symmetric about the $y$ axis, which lies along the  centerline. The
maximum Josephson critical current is $I_0$ for each of the
Josephson junctions, shown as small black squares. 
\begin{figure}
\includegraphics[width=8cm]{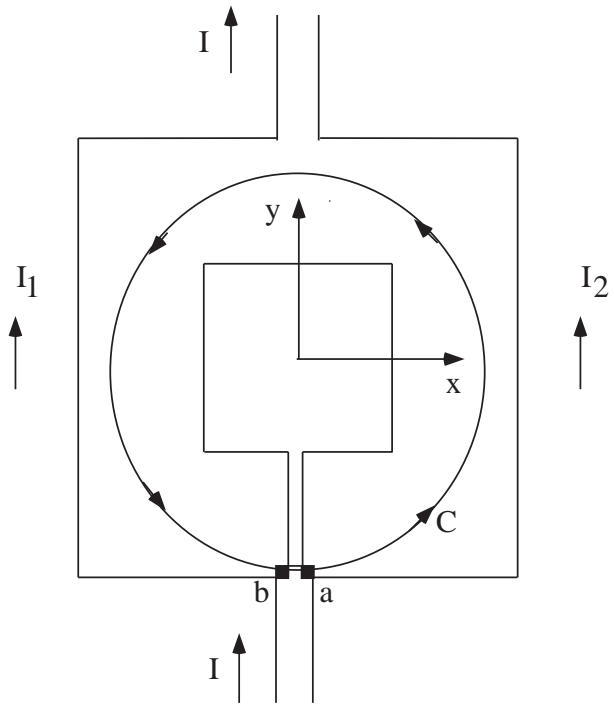}
\caption{
$I$ enters the main body of the SQUID from the counterelectrode
below through two Josephson junctions (small black squares, labeled
$a$ and $b$) and divides into the currents $I_1 = I/2 -I_{\rm d}$
and
$I_2=I/2+I_{\rm d}$ as shown.}
\label{fig1}
\end{figure} The currents up through the left and right sides of
the SQUID can be written as $I_1 = I/2 -I_{\rm d}$ and $I_2 = I/2
+I_{\rm d}$.  When the magnitude of
$I$, the total current through the SQUID, is less than the critical
current $I_{\rm c}$, the equations that determine $I=I_1+I_2$ and the
circulating current $I_{\rm d} = (I_2-I_1)/2$ can be derived using
a method similar to that used in Ref.\ \onlinecite{Orlando91}. We
begin by writing the local current density ${\bm j}$ in the
superconductors (i.e., the main body of the SQUID and the
counterelectrode) as\cite{London61}
\begin{equation}  
{\bm j} = -(1/\mu_0
\lambda^2) [{\bm A} + (\phi_0/2\pi)\nabla \gamma],
\label{j}
\end{equation}
   where  $\bm A$ is the vector potential, $\phi_0 = h/2e$ is the
superconducting flux quantum, and
$\gamma$ is the phase of the order parameter. The quantity inside
the brackets, which is gauge-invariant, can be thought of as the
superfluid velocity expressed in units of vector potential.  From
the point of view of the Ginzburg-Landau theory, implicit in the
use of this London-equation approach  is  the assumption that the
applied fields and currents are so low that the magnitude of the
order parameter is not significantly reduced from its equilibrium
value in the absence of fields and currents.

To obtain the SQUID equations, we integrate the vector potential
around a contour $C$  that passes in a  counterclockwise direction
through both junctions, the main body of the SQUID, and the
counterelectrode as shown in Fig.\ 1 and write the result in two
ways. Since $\bm B = \nabla \times \bm A$, this integral yields, on
the one hand, the magnetic flux in the $z$ direction
\begin{equation}  
\int_S B_z(x,y) dS,
\label{Phiz}
\end{equation} where $S$ is the area surrounded by the contour $C$
and $B_z(x,y)$ is the $z$ component of the net magnetic induction
in the plane of the SQUID produced by the sum of a perpendicular
applied field $B_{\rm a }$ and the self-field $B_{sz}(x,y)$ generated
via the Biot-Savart law by the supercurrent density ${\bm j}(x,y)$. On
the other hand, for those portions of the contour lying in the
superconductors, we  eliminate the line integrals of $\bm A$ in
favor of  line integrals of $\bm j$ using Eq.\ (\ref{j}).  We then
express the line integrals of
$\nabla \gamma$ in terms of the values of $\gamma$ at the junctions.
Equating the two expressions for the line integral of $\bm A$, we
find that the {\it effective flux} $\Phi$ in the $z$ direction through
the SQUID is given by
\begin{equation}  
\Phi = (\phi_0/2\pi)(\phi_1 -\phi_2),
\label{Fphi}
\end{equation} where
\begin{equation}  
\Phi = \int_S B_z(x,y) dS + \mu_0 \lambda^2 \!\int_C {\bm j}
\cdot {\bm {dl}},
\label{Fj}
\end{equation} with the integration contour $C$ now passing through
both superconductors but excluding the junction barriers. These equations
are equivalent to Eq.\ (8.67) in Ref.\
\onlinecite{Orlando91}. The gauge-invariant phase differences
across the junctions
$b$ and
$a$ are, respectively,\cite{Josephson69}
\begin{eqnarray}  
\phi_1 = \gamma_{bc}-\gamma_{bs} -(2\pi/\phi_0)\int_{bc}^{bs} \!
{\bm A} \cdot {\bm {dl}},\\
\label{phi1}
\phi_2 = \gamma_{ac}-\gamma_{as} -(2\pi/\phi_0)\int_{ac}^{as} \!
{\bm A} \cdot {\bm {dl}},
\label{phi2}
\end{eqnarray} where $bc$ labels a point on the counterelectrode
side of the junction $b$ and $bs$ labels the point directly across
the insulator in the SQUID washer, and $ac$ and $as$ label
corresponding points for junction $a$.  According to the Josephson
equations,\cite{Josephson69}  the junction supercurrents are
$I_1 = I_0 \sin\phi_1$ and $I_2 = I_0 \sin \phi_2$.
In the above derivation we have assumed that the linear dimensions of the
Josephson junctions are much less than the Josephson penetration depth
$\lambda_J$,\cite{Josephson69} and that the applied fields are
sufficiently small that the Josephson current densities and
gauge-invariant phase differences are very nearly constant across the
junction areas.

The magnetic moment $\bm m = m \hat z$ generated by the currents in the
SQUID is\cite{Jackson62}
\begin{equation} 
\bm m = \frac{1}{2}\int \bm r \times \bm j d^3r.
\label{m}
\end{equation}

It can be shown with the help of the London fluxoid quantization
condition,\cite{London61}
\begin{equation}  
\int_{S'} \! B_z(x,y) dS + \mu_0 \lambda^2 \! \int_{C'}\! {\bm j}
\cdot {\bm {dl}} = n \phi_0,
\label{Londonj}
\end{equation} where $n$ is an integer and $C'$ is a closed contour
that surrounds an area $S'$ within the body of the SQUID, that if
there are no vortices present (i.e., when $n=0$), the expression
for $\Phi$ in Eq.\ (\ref{Fj}) is independent of the choice of contour
$C$. Any  convenient path can be chosen for $C$, provided only that
the path remains in the superconducting material in the body of the
SQUID and the counterelectrode.  On the other hand, when there are
vortices in the main body of the SQUID, the quantity $\Phi$ increases
by
$\phi_0$ each time the contour $C$ is moved from a path inside the
vortex axis to one enclosing the vortex axis.  Thus, without
specifying the precise contour $C$,
$\Phi$ is determined only modulo $\phi_0$.  However, this is of no
physical consequence, because the gauge-invariant phases $\phi_1$
and $\phi_2$, which also enter Eq.\ (\ref{Fphi}), are also
determined only modulo $2 \pi$.  The final equations determining
the currents $I$ and $I_{\rm d}$ are independent of the choice of
contour $C$ and remain valid even when vortices are present in the
main body of the SQUID.

When the thickness $d$ of the SQUID is much larger than $\lambda$,
the contours $C$ and $C'$ can be chosen to be at the midpoint of the
thickness, where $\bm j$ is exponentially small, such that the line
integrals of
$\bm j$ can be neglected.  
The resulting equations are then the familiar ones found in many
reference books, such as Refs.\  
\onlinecite{Orlando91,deBruyn70,Solymar72,deBruyn77,VanDuzer81,Clarke90,
Tinkham96}.
However,
we are interested here in the case for which $d < \lambda$,
such that the fields
and currents are governed by the two-dimensional screening length or Pearl
length~\cite{Pearl64}
$\Lambda=\lambda^2/d$.  
The term in Eq.\ (\ref{Fj}) involving  $\bm j$ then must be carefully
accounted for.  For this case,
$\bm j$ is very nearly constant over the thickness and it is more
convenient to deal with the sheet-current density $\bm J(x,y) =  {\bm
j}d$, such that Eqs. (\ref{Fj}) and (\ref{Londonj}) take the
form\cite{foot4}
\begin{equation}  
\Phi = \int_S B_z(x,y) dS+ \mu_0 \Lambda \int_C {\bm J}
\cdot {\bm {dl}}
\label{FJ}
\end{equation} and
\begin{equation}  
\int_{S'} B_z(x,y) dS + \mu_0 \Lambda \int_{C'} {\bm J}
\cdot {\bm {dl}} = n \phi_0.
\label{LondonJ}
\end{equation}

For the general case when the SQUID is subject to a perpendicular
applied magnetic induction $B_{\rm a }$, carries a current $I$
unequally divided between the two arms,
$I_1 = I/2-I_{\rm d}$ and  $I_2 = I/2+I_{\rm d}$, where the
circulating current\cite{foot1} is $I_{\rm d} = (I_2-I_1)/2$, and
contains a vortex at the position ${\bm r}_{\rm v}$ in the body of
the SQUID, the {\it effective flux} $\Phi$ in the $z$ direction can be
written as the sum of four independent contributions:
\begin{equation}  
\Phi =\Phi_I + \Phi_{\rm d} + \Phi_{\rm f} +  \Phi_{\rm v},
\label{F}
\end{equation}

The first term on the right-hand side of Eq.\ (\ref{F}) is that
which would be produced by equal currents $I/2$ in the $y$
direction on the left and right sides of the SQUID shown in Fig.\ 1:
\begin{equation}  
\Phi_I = \int_{S} B_I(x,y) dS + \mu_0
\Lambda \int_C {\bm J}_I
\cdot {\bm {dl}},
\label{FI}
\end{equation} where  $B_I(x,y)$ is the $z$ component of the
self-field generated via the Biot-Savart law by the  sheet-current
density
${\bm J}_I(x,y)$, subject to the condition that the same current
$I/2$ flows through the two contacts $a$ and $b$.  For a symmetric
SQUID, $J_I(x,y)$, the $y$ component of ${\bm J}_I(x,y)$, is then an
even function of
$x$, and
$J_{I}(x,y)$ and $B_I(x,y)$ are odd functions of $x$.  As a
result, both terms on the right-hand side of Eq.\ (\ref{FI}) vanish
by symmetry, and $\Phi_I = 0$. Since $\nabla \cdot  {\bm J}_{I} = 0$
except at the contacts $a$ and $b$, we may write
${\bm J}_{I} = -(I/2) \nabla \times {\bm G}_I$, where
${\bm G}_I = \hat z G_I,$ such that
${\bm J}_I(x,y) = (I/2) \hat z \times \nabla G_I(x,y)$. The contours
of the scalar stream function
$G_I(x,y) = $ const correspond to streamlines of ${\bm J}_I(x,y)$,
and we may choose $G_I = 0$ for points ${\bm r}_i = (x_i,y_i)$ all
along the inner edges of the superconductors and
$G_I = 1$ for points ${\bm r}_o=(x_o,y_o)$ all along the outer right
edges and $G_I = -1$ for points ${\bm r}_o=(x_o,y_o)$ all along the
outer left edges.

The second term on the right-hand side of Eq.\ (\ref{F}) is due to
the circulating current\cite{foot1} $I_{\rm d} = (I_2-I_1)/2$  in
the counterclockwise direction when unequal currents flow in the two
sides of the SQUID shown in Fig.\ 1:
\begin{equation}  
\Phi_{\rm d} = \int_{S} B_{\rm d}(x,y) dS +
\mu_0 \Lambda \int_C {\bm J}_{{\rm d}}
\cdot {\bm {dl}},
\label{Fcirc}
\end{equation} where  $B_{\rm d}(x,y)$ is the $z$ component of
the self-field generated via the Biot-Savart law by the circulating
sheet-current density
${\bm J}_{{\rm d}}(x,y)$ when a current $I_{{\rm d}}$  flows through
contact $a$ from the counterelectrode into the body of the SQUID,
passes around the central hole, and flows through contact
$b$ back into the counterelectrode.  The magnetic moment
$m_d$ generated by the circulating current is proportional to $I_d$, as
can be seen from Eq.\ (\ref{m}). Since $\nabla \cdot  {\bm
J}_{{\rm d}} = 0$ except  at the contacts $a$ and
$b$, we may write
${\bm J}_{{\rm d}} = -I_{{\rm d}} \nabla \times {\bm G}_{{\rm d}}$,
where ${\bm G}_{{\rm d}} = \hat z G_{{\rm d}}$, such that
${\bm J}_{{\rm d}}(x,y) = I_{{\rm d}} \hat z \times \nabla G_{{\rm
d}}(x,y)$. The contours of the scalar stream function
$G_{{\rm d}}(x,y) = $ const correspond to streamlines of
${\bm J}_{{\rm d}}(x,y)$, and we may choose $G_{{\rm d}} = 0$ for
points ${\bm r}_i = (x_i,y_i)$ all along the inner edges of the
superconductors and
$G_{{\rm d}} = 1$ for points ${\bm r}_o=(x_o,y_o)$ all along the
outer edges. Once a numerical result for $\Phi_{{\rm d}}$ is found,
the result can be used to determine the inductance $L$ of the SQUID
via $L = \Phi_{{\rm d}}/I_{{\rm d}}$, as was done for a circular ring
in Ref.\ \onlinecite{Brandt04}. The resulting expression for $L$ is
the sum of the geometric and kinetic inductances.

The third term on the right-hand side is a {\it flux-focusing} term
due to the applied field:
\begin{equation}  
\Phi_{\rm f} = \int_{S} B_{\rm f}(x,y) dS +
\mu_0 \Lambda \int_C {\bm J}_{\rm f}
\cdot {\bm {dl}},
\label{Fa}
\end{equation} where  $B_{\rm f}(x,y)$ is the $z$ component of
the net magnetic induction in the plane of the SQUID produced by
the sum of a perpendicular applied field $B_{\rm a }$ and the $z$
component of the self-field $B_{s\rm f}(x,y)$ generated via the
Biot-Savart law by the  sheet-current density
${\bm J}_{\rm f}(x,y)$ induced in response to $B_{\rm a }$, subject
to the condition that no current flows through the junctions $a$ and
$b$.  
In other words, the desired fields are those that would appear in
response to $B_{\rm a }$ if the junctions  $a$ and
$b$ were open-circuited. Since
$\nabla \cdot  {\bm J}_{\rm f} = 0$, we may write
${\bm J}_{\rm f} = - \nabla \times {\bm G}_{\rm f}$, where
${\bm G}_{\rm f} = \hat z G_{\rm f}$, such that
${\bm J}_{\rm f}(x,y) = \hat z \times \nabla G_{\rm f}(x,y)$. The
contours of the scalar stream function $G_{\rm f}(x,y) = $ const
correspond to streamlines of
${\bm J}_{\rm f}(x,y)$, and we may chose $G_{\rm f} = 0$ for all
points $(x,y)$ along the inner and outer edges of the
superconductor. Once a numerical result for $\Phi_{\rm f}$ is found,
the result can be used to determine the  effective
area\cite{Ketchen85} of the SQUID's central hole, $A_{\rm eff} =
\Phi_{\rm f}/B_{\rm a }$, as was done for a circular ring in Ref.\
\onlinecite{Brandt04}.

To prove that the effective area is also given by $A_{\rm eff}
= m_d/I_d$,\cite{foot2} we consider the
electromagnetic energy cross term $E_{{\rm fd}} = 
\int (\bm B_{{\rm f}}\cdot \bm B_{{\rm d}}/\mu_0
+\mu_0 \lambda^2 \bm j_{{\rm f}}\cdot \bm j_{{\rm d}}) d^3r$, where the
integral extends over all space.  
Here, $\bm B_{\rm f}(\bm r) = \bm B_{\rm a}(\bm r) + \bm B_{s\rm f}(\bm r)
=
\nabla \times \bm A_{\rm f}(\bm r),$  where $\bm j_{\rm a}(\bm r) = \nabla
\times
\bm B_{\rm a}(\bm r)/\mu_0$ is the current density in the distant coil
that produces a nearly uniform field $B_a$ in the vicinity of the SQUID,
$\bm j_{\rm f} = \nabla \times \bm
B_{s\rm f}/\mu_0$ is the induced current density in the SQUID, and  $\bm
B_{s\rm f}$ is the corresponding self-field under the conditions of flux
focusing, i.e., when $\bm
j_{\rm f} = 0$ through the junctions. 
Also, $\bm B_{\rm d} = \nabla
\times \bm A_{\rm d}$ is the dipole-like field distribution
generated by the circulating current $I_{\rm d}$ with density $\bm
j_{{\rm d}}$ in the SQUID; at large distances from the
SQUID\cite{Jackson62} $\bm A_{\rm d} =
\mu_0 \bm m_{\rm d} \times \bm r /4 \pi r^3$.
We evaluate  $E_{{\rm fd}}$ in two ways, making
use of the vector identities
$\nabla
\cdot (\bm A \times \bm B) = \bm B \cdot (\nabla \times \bm A) 
-\bm A \cdot(\nabla \times \bm B)$ and $\nabla \cdot(\gamma \bm j) =
\gamma \nabla \cdot \bm j +\nabla \gamma  \cdot \bm j$, and applying the
divergence theorem, first with $\bm A = \bm A_{\rm d}$,
$\bm B =
\bm B_{\rm f}$, $\gamma = \gamma_{\rm d}$, and $\bm j = \bm j_{\rm f}$,
from which we obtain $E_{\rm {fd}} = B_{\rm a} m_{\rm d}$, and then with
$\bm A = \bm A_{\rm f}$,
$\bm B =
\bm B_{\rm d}$, $\gamma = \gamma_{\rm f}$, and $\bm j = \bm j_{\rm d}$,
from which we obtain  $E_{\rm {fd}} = \Phi_{\rm f} I_{\rm d}$ with the help
of Eq.\ (\ref{Fphi}).  Since $ \Phi_{\rm f} = B_{\rm a} A_{\rm {eff}}$,
the effective area obeys
$A_{\rm eff} = m_d/I_d$.

The fourth term on the right-hand side of Eq.\ (\ref{F}) is due to
a vortex at position ${\bm r}_{\rm v} = \hat x x_{\rm v} + \hat y
y_{\rm v}$ in the body
of the SQUID:
\begin{equation}  
\Phi_{\rm v}({\bm r}_{\rm v}) = \int_{S} B_{\rm v}(x,y)
dS + \mu_0 \Lambda
\int_C {\bm J}_{\rm v} \cdot {\bm {dl}},
\label{Fv}
\end{equation} where  $B_{\rm v}(x,y)$ is the $z$ component of the
self-field  generated by the vortex's  sheet-current density
${\bm J}_{\rm v}(x,y)$ via the Biot-Savart law when no current flows
through the junctions $a$ and $b$. 
The desired fields are those that would appear in
response to the vortex if the junctions  $a$ and
$b$ were open-circuited.  Since $\nabla
\cdot  {\bm J}_{\rm v} = 0$, it is possible to express ${\bm J}_{\rm
v}(x,y)$ in
terms of a scalar stream function, as we did for ${\bm J}_{\rm f}(x,y)$
and ${\bm J}_{{\rm d}}(x,y)$.  However, as shown below, it is possible
to use energy arguments to express
$\Phi_{\rm v}(x,y)$ in terms of the stream function $G_{{\rm
d}}(x,y)$.\cite{foot3}

To obtain $\Phi_{\rm v}({\bm r})$ when a vortex is at the position
${\bm r} =
\hat x x + \hat y y$, imagine disconnecting the counterelectrode in
Fig.\ 1 and attaching leads from a power supply to the contacts $a$
and $b$.  The power supply provides a constant current $I_{{\rm
d}}$ in the counterclockwise direction, and the sheet-current
distribution through the body of the SQUID is given by
${\bm J}_{{\rm d}}(x,y) = I_{{\rm d}} \hat z \times
\nabla G_{{\rm d}}(x,y)$, as discussed above. We also imagine
attaching leads from a high-impedance voltmeter to the contacts $a$
and $b$.  If the vortex moves, the effective flux $\Phi_{\rm v}$
changes with time, and the voltage read by the voltmeter will
be\cite{Clem70}
$V_{ab} = d\Phi_{\rm v}/dt$.  The power  delivered by the power supply
can be expressed in terms of the Lorentz force on the vortex,
${\bm J}_{{\rm d}} \times \hat z \phi_0 = I_{{\rm d}} \phi_0
\nabla G_{{\rm d}}$; i.e., the rate at which work is  done on the
moving vortex is
$I_{{\rm d}} \phi_0 \nabla G_{{\rm d}} \cdot d{\bm r}/dt$. Equating
this to the power
$P = I_{{\rm d}}d\Phi_{\rm v}/dt = I_{{\rm d}} \nabla \Phi_{\rm v} \cdot
d{\bm r}/dt$ delivered by the power supply to maintain constant
current, we obtain the equation
$\nabla \Phi_{\rm v}(\bm r) = \phi_0 \nabla G_{{\rm d}}(\bm r)$.  Thus
$\Phi_{\rm v}(\bm r) = \phi_0 G_{{\rm d}}(\bm r) + $const, where the
constant can have one of two possible values depending upon whether
the integration contour
$C$ is chosen to run inside or outside the  vortex axis at
${\bm r} = \hat x x + \hat y y$.  Choosing  $C$ to run around the
outer boundary of the SQUID, we obtain
\begin{equation}  
\Phi_{\rm v}({\bm r})=\phi_0G_{{\rm d}}({\bm
r}).
\label{FvG}
\end{equation} Since $G_{{\rm d}}({\bm r}_o) = 1$  for points
${\bm r} = {\bm r}_o$ on the outer edges of the SQUID and
$G_{{\rm d}}({\bm r}_i) = 0$ for points ${\bm r} = {\bm r}_i$ on
the inner edges (at the perimeter of the central hole or along the
edges of the slit), we have
$\Phi_{\rm v}({\bm r}_o) = \phi_0$ and $\Phi_{\rm v}({\bm r}_i) = 0$.
The
derivation of Eq.\ (16 ) implicitly assumes that the vortex-core radius
is much smaller than the linear dimensions of the SQUID.

We now return to the problem of how to find the currents $I$ and
$I_{{\rm d}}$ in the SQUID, as well as the critical current
$I_{\rm c}$.
As discussed above, we have $\Phi_I = 0$ for a symmetric SQUID.  For
simplicity, we assume first that there are no vortices in the body
of the SQUID, such that
$\Phi_{\rm v} = 0$ and $\Phi = \Phi_{\rm f} + \Phi_{{\rm d}}$ in Eq.\
(\ref{Fphi}), where
$\Phi_{{\rm d}} = LI_{{\rm d}}$. From  the sum and the difference of
$I_1$ and $I_2$ we obtain
\begin{eqnarray}   
I = 2 I_0 \cos\Big(\frac{\pi \Phi_{\rm f}}{\phi_0} +\frac{\pi
 L I_{\rm d}} {\phi_0}\Big)\sin \bar\phi, \label{I}\\
 I_{{\rm d}} = - I_0 \sin\Big(\frac{\pi \Phi_{\rm f}}{\phi_0}
+\frac{\pi L I_{\rm d}}{\phi_0}\Big)\cos \bar\phi,
\label{Icirc}
\end{eqnarray} where $\bar \phi = (\phi_1 + \phi_2)/2$ is
determined experimentally by how much current is applied to the
SQUID.  When $\bar \phi = 0$, the current $I$ is zero.  As $\bar
\phi$ increases, the magnitude of $I$ increases and reaches its
maximum value $I_{\rm c}$ at a value of  $\bar \phi$ that must be
determined by numerically solving Eqs.\ (\ref{I}) and
(\ref{Icirc}).  A simple solution is obtained for arbitrary
$\Phi_{\rm f}$ only in the limit $\pi L I_0/\phi_0 \rightarrow 0$,
for which $I = I_{\rm c} = 2I_0|\cos (\pi \Phi_{\rm f}/\phi_0)|$ and
$I_{{\rm d}} = 0$ at the critical
current. For values of  $\pi L I_0/\phi_0$ of order unity, as is the
case for practical SQUIDs, one may obtain $I_{\rm c}$ for any value
of $\Phi_{\rm f}$ by solving Eq.\ (\ref{Icirc}) self-consistently for
$I_{{\rm d}}$ for a series of  values of $\bar \phi$ and by
substituting the results into Eq.\ (\ref{I}) to determine which
value of $\bar \phi$ maximizes $I$.  Equations (\ref{I}) and
(\ref{Icirc}) have been solved numerically by de Bruyn Ouboter and
de Waele,\cite{deBruyn70} (some of their results are also shown by
Orlando and Delin\cite{Orlando91}), who showed that at $I_{\rm c}$
\begin{eqnarray}   
I_{\rm c}(\Phi_{\rm f}) &=& I_{\rm c}(\Phi_{\rm f} + n
\phi_0) = I_{\rm c}(-\Phi_{\rm f}),
\label{Icsymm} \\  
I_{{\rm d}}(\Phi_{\rm f}) &=& I_{{\rm
d}}(\Phi_{\rm f} + n \phi_0) =
   -I_{{\rm d}}(-\Phi_{\rm f}),
\label{isymm} \\   
I_1(\Phi_{\rm f}) &=& I_1(\Phi_{\rm f} + n
\phi_0) = I_2(-\Phi_{\rm f}),
\label{I1symm} \\  
I_2(\Phi_{\rm f}) &=& I_2(\Phi_{\rm f} + n
\phi_0) = I_1(-\Phi_{\rm f}),
\label{I2symm}
\end{eqnarray} where $n$ is an integer.  Hence all the physics is
revealed by displaying $I_{\rm c}(\Phi_{\rm f})$ over the interval
$0 \le
\Phi_{\rm f} \le \phi_0/2$, as shown in Fig.\ 2.
\begin{figure}
\includegraphics[width=8cm]{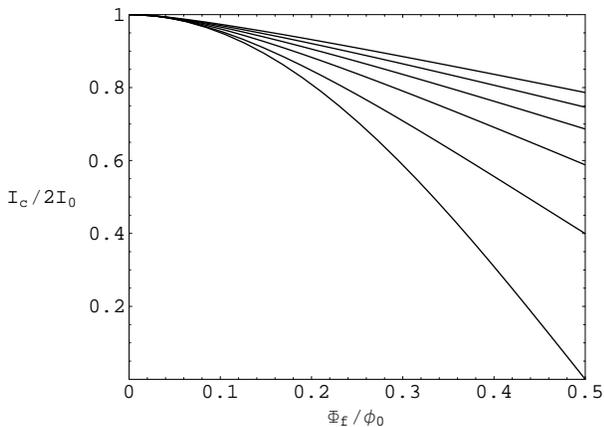}
\caption{%
$I_{\rm c}/2I_0$ vs $\Phi_{\rm f}/\phi_0$, calculated from
Eqs.\ (\ref{I}) and (\ref{Icirc}), for $\pi L I_0/\phi_0$ = 0,
 1, 2, 3, 4, and 5 (bottom to top).}
\label{fig2}
\end{figure}

When a vortex is present, Eqs.\ (\ref{I}) and (\ref{Icirc}) still
hold, except that $\Phi_{\rm f}$ in these equations is replaced by the
sum
$\Phi_{\rm f} + \Phi_{\rm v}$. Thermally agitated motion of vortices in
the body of the SQUID can produce flux noise via the term
$\Phi_{\rm v}({\bm r}_{\rm v})$ and the time dependence of the vortex
position
${\bm r}_{\rm v}.$ 
From  Eq.\
(\ref{FvG}) we see that the sensitivity of $I_c$ to vortex-position noise
is proportional to the magnitude of $\nabla \Phi_{\rm v}({\bm r})=\phi_0
\nabla  G_{{\rm d}}= {\bm J}_{{\rm d}} \times \hat z \phi_0 /I_{{\rm d}}.$
Thus
$I_{\rm c}$ is most sensitive to vortex-position noise when the
vortices
are close to the inner or outer edges of the SQUID, where 
the magnitude of ${\bm J}_{{\rm d}}$ is largest. 
These equations
provide more accurate results for the vortex-position
sensitivity than the approximations given in Refs.\
\onlinecite{Ferrari91} and \onlinecite{Sun94}.

So far, we have investigated how the general equations governing the
behavior of a dc SQUID are altered when the contributions arising
from  line integrals of the current density are included.  As we
have shown in Ref.\ \onlinecite{Brandt04}, these additional
contributions are important when the Pearl length
$\Lambda$ is an appreciable fraction of the linear dimensions of the
SQUID.  We have found that the basic SQUID equations, Eqs.\
(\ref{I}) and (\ref{Icirc}), are unaltered, except that the
magnetic flux (sometimes called
$\Phi_{\rm ext}$\cite{Orlando91}) generated in the SQUID's central
hole by the externally applied field 
in the absence of a vortex is replaced by the {\it effective
flux} $\Phi_{\rm f}$, given in Eq.\ (\ref{Fa}).  Similarly, we
have shown that the total inductance $L$ of the SQUID has contributions
both from the magnetic induction (geometric inductance) and the associated
supercurrent (kinetic inductance). We also have shown in principle how to
calculate the effect of the return flux from a vortex at position
$\bm r_{\rm v}$  in the body of the SQUID, and we have found that the
effective flux arising from the vortex is $\Phi_{\rm v}(\bm r_{\rm
v})$, given in Eqs.\ (\ref{Fv}) and (\ref{FvG}).  To demonstrate that all
the above quantities can be calculated numerically
for arbitrary values of $\Lambda$, we next examine the behavior of a model
SQUID as decribed in Sec.\ III. 
\begin{figure}
\includegraphics[width=6cm]{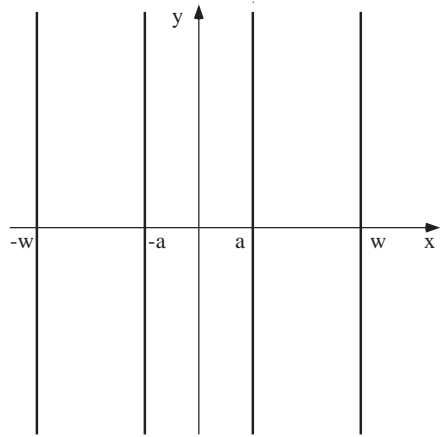}
\caption{%
Sketch of central portion of the long SQUID considered in Sec.\ III.}
\label{fig3}
\end{figure}

\section{Long SQUID in a
perpendicular magnetic field}

Here we consider a long SQUID whose thickness $d$ is less than the
London penetration depth $\lambda$ and  whose topology is like that
of Fig.\ 1 but which is stretched to a large length
$l$ in the $y$ direction, as sketched in Fig.~3.  SQUIDs of similar
geometry have been investigated experimentally in Refs.\
\onlinecite{Pegrum99,Eulenburg99,Kuriki99,Mitchell02,Mitchell03}. We treat
here the case for which the length $l$ is much larger than the
width $2w$ of the body of the SQUID, and we focus on the current
and field distributions in and near the left ($-w < x < -a$) and
right ($a < x < w$) arms and near the center of the SQUID, where to
a good approximation the current density $j_y$ is uniform across
the thickness and depends only upon $x$, and the magnetic induction
$\bm B = \nabla \times \bm A$ depends only upon $x$ and
$z$. In the equations that follow, we deal with the sheet-current
density, whose component in the $y$ direction is $J_y(x) = j_y(x)
d$.

The self-field magnetic induction generated by $J_y(x)$ is  $\bm
B_J(x,z) = \nabla \times \bm A_J(x,z)$, where $A_J(x,z)$, the $y$
component of the vector potential obtained from Ampere's law, is
\begin{equation}  
A_J(x,z)= \frac{\mu_0}{2 \pi}\int \!
J_y(x')
\ln \frac{C}{\sqrt{(x-x')^2 + z^2}}  \, dx'.
\label{Aygeneral}
\end{equation} The integration here and in the following equations
is carried out only over the strips, and $C$ is a constant with
dimensions of length remaining to be determined. In the presence of
a perpendicular applied field $\bm B_{\rm a } = \hat z B_{\rm a } =
\nabla \times \bm A_{\rm f}$, the total vector potential is $\bm A
= \bm A_J +
\bm A_{\rm f}$, where $\bm A_{\rm f} = \hat y B_{\rm a } x.$

\subsection{Formal solutions}

We  now use the approach of Ref.\ \onlinecite{Brandt01} to calculate the
in-plane magnetic-induction and sheet-current distributions appearing in
Eqs.\ (\ref{F})-(\ref{Fa}) in Sec.\ II. For all of these contributions we
shall take into account the in-plane ($z=0$) self-field contribution
$A_J(x) = A_J(x,0)$ to the $y$ component of the vector potential, where
\begin{equation}  
A_J(x) = \frac{\mu_0}{2 \pi}\int J_{y}(x')
\ln \frac{C}{|x-x'|}  \, dx'.
\label{AJy}
\end{equation}

We first examine the {\it equal-current case} and consider the
contributions
$B_I(x)$, the
$z$ component of
${\bm B}_I(x)$, and
$J_I(x)$, the $y$ component of ${\bm
J}_I(x)$, due to equal currents $I/2$ in the left and right sides of the
SQUID. Since $J_I(-x) = J_I(x)$, the corresponding $y$ component of
the vector potential is also a symmetric function of $x$: $A_{I}(-x) =
A_{I}(x)$, where the subcripts $I$ refer to the  equal-current case.
There are no flux quanta between the strips $(\Phi_I = 0$), and the second
term in the brackets on the right-hand side of  Eq.\ (\ref{j}) vanishes;
$(\phi_0/2 \pi) \nabla \gamma = 0$.  However, the constant
$C$ must be chosen such that $J_I(x) = - A_{I}(x)/\mu_0 \Lambda$
in the superconductor. Combining this equation with Eq.\
(\ref{AJy}), making use of the symmetry $J_I(-x) = J_I(x)$,
and noting that $I = 2\int_a^w J_I(x)dx$, we obtain
\begin{equation}  
\frac{I}{2\pi}\ln\frac{b}{C} =\! \int_a^w \!
\!\Big[\frac{1}{2\pi}
\ln\frac{b^2}{|x^2 \!-\! x^{'2}|}+\Lambda \delta(x \!-\! x')
\Big]J_I(x')dx'
\label{C}
\end{equation} for $a < x < w$. Here $b$ can be chosen to be any
convenient length, such as the length
$l$ or $w$, but not $C$.  We now define the inverse integral kernel
$K^{\rm sy}(x,x')$ for the symmetric-current case via
\begin{eqnarray}  
\int_a^w \! \!\!K^{\rm sy}(x,x'')\Big[\frac{1}{2\pi} \ln
\frac{b^2}{|x^{''2}\!-x^{'2}|}+\Lambda \delta(x''\!-x')\Big]dx''
\nonumber \\ = \delta(x-x').~
\label{Ksy}
\end{eqnarray} Applying this kernel to Eq.\ (\ref{C}), we obtain
\begin{equation}  
J_I(x) =
\frac{I}{2\pi}\ln\Big(\frac{b}{C}\Big) \int_a^w \!
\!K^{\rm sy}(x,x') dx'.
\label{JIyC}
\end{equation} Since $I = 2 \int_a^wJ_I(x)dx$, we find
\begin{equation}  
\ln\Big(\frac{b}{C}\Big)= \pi\Big/ \! \!\int_a^w  \! \! \!
\!\int_a^w
\! \!K^{\rm sy}(x,x') dx dx',
\label{lnbC}
\end{equation} such that
\begin{equation}  
  J_I(x) = \frac{I}{2}\! \int_a^w\! \!\! K^{\rm sy}(x,x')
dx' \!
  \Big/ \!\!\!\int_a^w \!\!\!\! \int_a^w \!\!\!K^{\rm
sy}(x',x'')dx'dx''.
\label{JIy}
\end{equation} For $a \le x \le w$, the stream function is
\begin{equation}  
G_I(x)=\frac{2}{ I} \int_a^x \!\!
J_I(x')dx',
\label{GI}
\end{equation} and for $-w \le x \le -a$, $G_I(-x)=-G_I(x)$. The
corresponding $z$ component of the magnetic induction $B_I(x)$ can be
obtained from the Biot-Savart law or, since $B_I(x) = dA_{I}(x)/dx,$ from
Eq.\ (\ref{AJy}). Note that $B_I(-x) = -B_I(x)$. Although the
kernel
$K^{\rm sy}(x,x')$ depends upon $b$, we find numerically that
$J_I(x)$ and $B_I(x)$ are independent of $b$.

We next examine the {\it circulating-current case} and consider the
contributions
$B_{\rm d}(x)$, the
$z$ component of
${\bm B}_{\rm d}(x)$, and
$J_{\rm d}(x)$, the $y$ component of ${\bm J}_{\rm d}(x)$, due to a
circulating current
$I_{{\rm d}}$; the current in the $y$ direction on the right side of the
SQUID is
$I_2 = I_{{\rm d}}$
  and that on the left side is $I_1 = -I_{{\rm d}}$.  The vector
potential is still given by Eq.\ (\ref{AJy}), except that we add
subscripts  ${\rm d}$. However, since
$J_{\rm d}(-x) = -J_{\rm d}(x)$, the vector potential is now
an antisymmetric function of $x$;
$A_{\rm d}(-x) = -A_{\rm d}(x)$.  The circulating current is
generated by the fluxoid
$\Phi_{{\rm d}}$ [see Eq.\ (\ref{Fcirc})] associated with a
nonvanishing gradient of the phase $\gamma$ around the loop.  The
second term inside the bracket on the right-hand side of Eq.\
(\ref{j}),
$(\phi_0/2\pi)\nabla \gamma$, is
$- \hat y \Phi_{{\rm d}}/2l$ for $a < x < w$ and
$\hat y \Phi_{{\rm d}}/2l$ for $-w < x < -a$.  Thus,
$J_{\rm d}(x) = - [A_{\rm d}(x)-\Phi_{{\rm d}}/2l]/\mu_0
\Lambda$ for $a<x<w$. Combining this equation with that for
$A_{\rm d}(x)$, noting that the inductance of the SQUID is $L=
\Phi_{{\rm d}}/I_{{\rm d}}$, and making use of the symmetry
$J_{\rm d}(-x) = -J_{\rm d}(x)$, we obtain
\begin{equation}  
\frac{L I_{{\rm d}}}{2 l} = \mu_0\! \int_a^w \!
\!\Big[\frac{1} {2\pi} \ln\Big|\frac{x+x'}{x-x'}\Big|+\Lambda
\delta(x-x')\Big]J_{\rm d}(x')dx'
\label{Fcirc/2l}
\end{equation} for $a < x < w$. We now define the inverse integral
kernel
$K^{\rm as}(x,x')$ for the asymmetric-current case via
\begin{eqnarray}  
\int_a^w \! \!K^{\rm as}(x,x'')\Big[\frac{1}{2\pi}
\ln\Big|\frac{x''+x'}{x''-x'}\Big|+\Lambda \delta(x''-x')\Big]dx''
\nonumber
\\=\delta(x-x').
\label{Kas}
\end{eqnarray} Applying this kernel to Eq.\ (\ref{Fcirc/2l}) and
noting that
$I_{{\rm d}} = \int_a^w J_{\rm d}(x)dx$, we obtain
\begin{equation}   
J_{\rm d}(x) =\alpha I_{{\rm d}}
\int_a^w \!
\!K^{\rm as}(x,x') dx'
\label{Jcircy}
\end{equation} and
\begin{equation}  
L=2\alpha \mu_0 l,
\label{L}
\end{equation}
   where
\begin{equation} 
\alpha= 1 \Big/ \! \!\int_a^w  \! \! \! \!\int_a^w
\! \!K^{\rm as}(x,x')dxdx'
\label{alpha}
\end{equation} is a dimensionless function of $a, b, w,$ and
$\Lambda$, which we calculate numerically in the next section. For
$a \le x
\le w$, the stream function is
\begin{equation}  
G_{\rm d}(x)=\frac{1}{ I_{\rm d}} \int_a^x \!J_{\rm d}(x')dx',
\label{Gcirc}
\end{equation} and for $-w \le x \le -a$, $G_{\rm d}(-x)=G_{\rm
d}(x)$.

In this formulation, as in Ref.\ \onlinecite{Brandt04},
$L = L_{\rm m} + L_{\rm k}$ is the total inductance.  The geometric
inductance contribution  $L_{\rm m}= 2E_{\rm m}/I_{\rm d}^2$, where
$E_{\rm m}=l\int_a^wJ_{\rm d}(x)A_{\rm d}(x)dx$ is the stored
magnetic energy, and the kinetic contribution
$L_{\rm k} = 2E_{\rm k}/I_{\rm d}^2$, where
$E_{\rm k}=\mu_0 \Lambda l \int_a^wJ^2_{{\rm d}}(x)dx$ is the
kinetic energy of the supercurrent, can be calculated  using Eq.\
(\ref{Jcircy}) from\cite{Brandt04,Meservey69,Yoshida92} 
\begin{eqnarray}  
L_{\rm m}\! &=&\! \frac{ \mu_0 l }{\pi
I^2_{\rm d}}  \!\int_a^w
\!\!\!\! \int_a^w \!\! \ln\Big|\frac{x+x'}{x-x'}\Big| \,
J_{\rm d}(x)J_{\rm d}(x') dx dx',~  \nonumber \\
\label{Lm}\\ L_{\rm k}\! & =&\! \frac{ 2\mu_0 l \Lambda}{
I^2_{\rm d}} \!\! \int_a^w  \! \! \! J^2_{{\rm d}}(x) dx
=\frac{2\mu_0l\Lambda}{(w-a)}\frac{< \!J_{\rm d}^2\!>}{<\!J_{\rm d}\!>^2},
\label{Lk}
\end{eqnarray}
where the brackets ($<>$) denote averages over the film width.  We can
show that
$L_{\rm m} +L_{\rm k} = L$ with the help of Eqs.\ (\ref{Kas}) and
(\ref{alpha}).

The $z$ component of the magnetic induction $B_{\rm d}(x)$ generated by
$J_{\rm d}(x)$ can be obtained from the Biot-Savart law, or,
since
$B_{\rm d}(x) = dA_{\rm d}(x)/dx,$ from Eq.\ (\ref{AJy}).
Note that $B_{\rm d}(-x) = B_{\rm d}(x)$.

When $l\gg w$, the magnetic moment in the $z$ direction generated
by the circulating current $I_{\rm d}$ is (to lowest order in $w/l$)
\begin{equation}  
m_{\rm d} = 2l\int_a^w \!\! x
J_{\rm d}(x) dx,
\label{mcirc}
\end{equation} where the factor 2 accounts for the
fact\cite{Brandt94} that the currents along the  $y$ direction and
those along the $x$ direction at the ends (U-turn) give exactly the
same contribution to
$m_{\rm d}$, even in the limit $l \rightarrow \infty$. To next
higher order in $w/l$ one has to replace $l$ in Eq.\ (\ref{mcirc})
by $l -(w-a)q$, where $(w-a)q/2$ is the distance of the center of
gravity of the $x$-component of the currents near each end from
this end. For a single strip of width
$w-a$, one has, e.g., $q=1/3$ for the Bean critical state (with
rectangular current stream lines) and $q=0.47$ for ideal screening
($\Lambda \ll w$).\cite{EHB1}

We next examine {\it flux focusing}. As discussed in Sec.\
II, to calculate the effective area of the slot, we need to
calculate the fields produced in response to a perpendicular applied
magnetic  induction $B_{\rm a}$, subject to the condition that no current
flows through either junction. Since this is equivalent to having both
junctions open-circuited, the problem reduces to finding the fields
produced in the vicinity of a pair of long superconducting strips
connected by a superconducting link at only one end, i.e., when
the slot of width $2a$ between the two strips is {\it open} at one end. 
However, the desired fields may be regarded as the superposition of the
solutions of two separate problems when the slot has {\it closed} ends:
(a) the fields generated in response to $B_{\rm a}$, when $\Phi=0$ (the {\it
zero-fluxoid case}) and a clockwise screening current flows around the
slot [second term on the right-hand side of Eq.\ (\ref{Jay}) below], and
(b) the fields generated in the absence of  
$B_{\rm a}$, when flux quanta in the amount of $\Phi_{\rm f}$ are in the
slot and a counterclockwise screening current flows around the slot
[first term on the right-hand side of Eq.\ (\ref{Jay})].
The desired flux-focusing solution is obtained by setting the net
circulating current equal to zero.

The equations describing the fields in the flux-focusing case are derived
as follows. The
$z$ component of the net magnetic induction
$B_{\rm f}(x)$ is the sum of  $B_{\rm a }$ and the self-field
$B_{s\rm f}(x)$ generated by $J_{\rm f}(x)$. The vector
potential $A_y(x)$ is the sum of $B_{\rm a } x$, which describes the
applied magnetic induction, and the self-field contribution given
by Eq.\ (\ref{AJy}) but with subscripts   f. Since
$J_{\rm f}(-x) = -J_{\rm f}(x)$, the vector potential is
again an antisymmetric function of $x$; $A_{\rm f}(-x) =
-A_{\rm f}(x)$. The fluxoid $\Phi_{\rm f}$ [see Eq.\ (\ref{Fa})]
contributes a nonvanishing gradient of the phase $\gamma$ around
the loop, such that the second term inside the brackets on the
right-hand side of Eq.\ (\ref{j}), $(\phi_0/2\pi)\nabla \gamma$, is
$-\hat y \Phi_{\rm f}/2l$ for $a < x < w$ and  $\hat y \Phi_{\rm f}/2l$
for $-w < x < -a$. Equation (\ref{j}) yields
$J_{\rm f}(x) = -[B_{\rm a } x + A_{\rm f}(x)-\Phi_{\rm f}/2l]/
\mu_0 \Lambda$  for
$a < x < w$. Combining this equation with that for $A_{\rm f}(x)$
[Eq.\ (\ref{AJy})], making use of the symmetry
$J_{\rm f}(-x) = -J_{\rm f}(x)$, and introducing the
effective area via
$\Phi_{\rm f} = B_{\rm a } A_{\rm eff}$ [see Sec.\ II], we obtain
\begin{flalign}  
&B_{\rm a } (A_{\rm eff} - 2lx)/2l = \nonumber
\\ &\mu_0 \! \int_a^w \! \!\Big[\frac{1}{2\pi}
\ln\Big|\frac{x+x'}{x-x'}\Big|+\Lambda
\delta(x-x')\Big] \, J_{\rm f}(x')dx'
\label{BaAeff}
\end{flalign} for $a < x < w$. We again use the inverse integral
kernel
$K^{\rm as}(x,x')$ for the asymmetric-current case [Eq.\
(\ref{Kas})] to obtain
\begin{equation}  
J_{\rm f}(x) = \frac{B_{\rm a }}{\mu_0}
\int_a^w \!
\!\big(\frac{A_{\rm eff}}{2l}-x'\,\big) K^{\rm as}(x,x') dx'.
\label{Jay}
\end{equation} The effective area of the SQUID $A_{\rm eff}$ is
found from the condition that the net  current around the loop is
zero [$\int_a^w J_{\rm f}(x)dx = 0$], which yields
\begin{equation} 
A_{\rm eff}= 2 \alpha l  \! \!\int_a^w
\! \! \! \!\int_a^w
\! \!\!x'K^{\rm as}(x,x')dxdx'.
\label{Aeff}
\end{equation} For $a \le x \le w$, the stream function is
\begin{equation}  
G_{\rm f}(x)=\int_a^x \! J_{\rm f}(x')dx',
\label{Ga}
\end{equation}
and for $-w \le x \le -a$, $G_{\rm f}(-x)=G_{\rm f}(x)$.
The spatial distribution of the resulting $z$ component of the in-plane
magnetic induction is given by
$B_{\rm f}(x) = B_{\rm a } + B_{s\rm f}(x)$, where
$B_{s\rm f}(x)$ can be obtained from the Biot-Savart law or by
substituting $J_{\rm f}(x)$ into Eq.\ (\ref{AJy}) and making
use of $B_{s\rm f}(x) = dA_{\rm f}(x)/dx$.  Note that
$B_{\rm f}(-x) = B_{\rm f}(x)$. The resulting magnetic moment
$m_{\rm f}$ in the $z$ direction can be calculated by replacing
$J_{\rm d}$  by
$J_{\rm f}$ in Eq.\ (\ref{mcirc}).

In the next section we also present numerical results for field and
current distributions in the {\it zero-fluxoid case}, in which $I =
0$,
$\Phi_{\rm v} = 0$, and the effective flux is zero: 
$\Phi = \Phi_{\rm d} + \Phi_{\rm
f} = 0$.  Such a case could be achieved by short-circuiting the Josephson
junctions in Fig.\ 1, cooling the device in zero field such that
initially $\Phi = 0$, and then applying a small perpendicular magnetic
induction
$B_{\rm a}$.  A circulating current $J(x)$, given by the
second term on the right-hand side of Eq.\ (\ref{Jay}), would
spontaneously arise in order to keep
$\Phi = 0,$ as in the Meissner state.

\subsection{Numerical solutions}

In the previous section we have presented formal solutions for the
sheet-current density  $J_y(x)$ in Eqs.\ (\ref{JIy}),
(\ref{Jcircy}), and (\ref{Jay}), which are expressed as integrals
involving the geometry-dependent inverse kernels $K^{\rm sy}(x,x')$
and  $K^{\rm as}(x,x')$. As in Ref.~\onlinecite{Brandt04} for thin
rings, these integrals are evaluated on a grid $x_i$ ($i=1, 2,
\dots, N$) spanning only the strip (but avoiding the edges, where
the integrand may have infinities), $a < |x_i| < w$, such that for
any function
$f(x)$ one has $\int_a^w f(x) dx =\sum_{i=1}^N w_i f(x_i)$.
Here $w_i$ are the weights, approximately equal to the local
spacing of the $x_i$; the weights obey $\sum_{i=1}^N w_i = w-a$. We
have chosen the grid such that the weights $w_i$ are narrower and
the grid points $x_i$ more closely spaced near the edges $a$ and
$w$, where $J_y(x)$ varies more rapidly. We have accomplished this
by choosing some appropriate continuous function $x(u)$ and an
auxiliary discrete variable $u_i \propto i - \frac{1}{2}$, such
that $w_i = x'(u_i) (u_2 -u_1)$. We can choose $x(u)$ such that its
derivative $x'(u)$ vanishes (or is reduced) at the strip edges to
give a denser grid there. By choosing an appropriate substitution
function $x(u)$ one can make the numerical error of this
integration method arbitrarily small, decreasing rapidly with any
desired negative power of the grid number $N$, e.g., $N^{-2}$ or
$N^{-3}$.

For the {\em equal-current case},  Eq.\ (\ref{C}) becomes
\begin{equation}  
\frac{I}{2\pi}\ln\frac{b}{C} = \sum_{j=1}^N (w_j Q^{\rm sy}_{ij}
+\Lambda \delta_{ij})J_I(x_j)
\label{Ci}
\end{equation} where $\delta_{ij}=0$ for $i\ne j$, $\delta_{ii}=1$,
and
\begin{eqnarray}  
   Q^{\rm sy}_{ij} &=&\frac{1}{2\pi} \ln{\frac{b^2 }{
      |x_i^2 -x_j^2|}}, ~~ i\ne j \,, \nonumber \\
   Q^{\rm sy}_{ii} &=&{\frac{1}{ 2\pi}} \ln{\frac{\pi b^2}{ x_i
w_i}}.
\label{Qsyij}
     \end{eqnarray} 
The optimum choice of the diagonal term $Q^{\rm
sy}_{ii}$ (i.e., with
$|x_i^2 - x_j^2|$ replaced by $x_iw_i/\pi$ for $i=j$, which reduces
the numerical error from order $N^{-1}$ to $N^{-2}$ or higher
depending on the grid) is discussed in Eq.\ (3.12) of Ref.\
\onlinecite{Brandt94} for strips and in Eq.\ (18) of Ref.\
\onlinecite{Brandt04} for disks and rings. The superscript (sy) is
a reminder that this is for a symmetric current distribution
[$J_I(-x) = J_I(x)$]. Defining $K^{\rm sy}_{ij}=(w_j Q^{\rm
sy}_{ij}+\Lambda
\delta_{ij})^{-1}$, such that
\begin{figure}
\includegraphics[width=8cm]{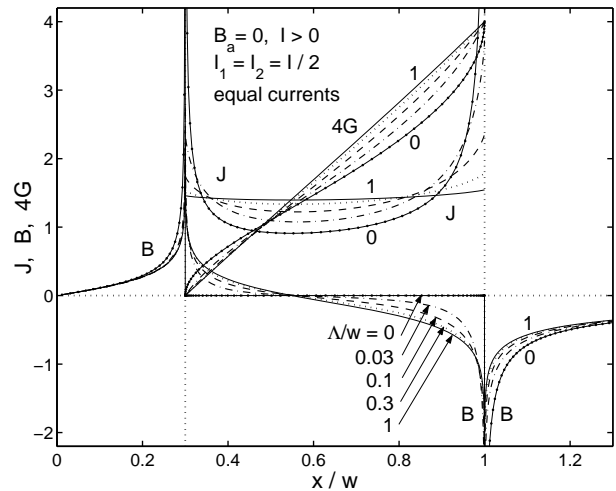}
\caption{%
  Profiles of the sheet current $J_I(x)$, Eq.~(\ref{JIyifinal}),
stream function $G_I(x)$, Eq.~(\ref{GIi}), and magnetic induction
$B_I(x)$ for the {\it equal-current case} ($B_a=0$, $I_1=I_2=I/2>0$).
Shown are the examples $a/w=0.3$ with $\Lambda/w =$ 0 (solid lines
with dots), 0.03 (dot-dashed lines), 0.1 (dashed lines), 0.3
(dotted lines), and 1 (solid lines). Here $B/\mu_0$ and $J$ are
in units $I_2 /w$ and $G$ in units $I_2$.}
\label{fig4}
\end{figure}
\begin{figure}
\includegraphics[width=8cm]{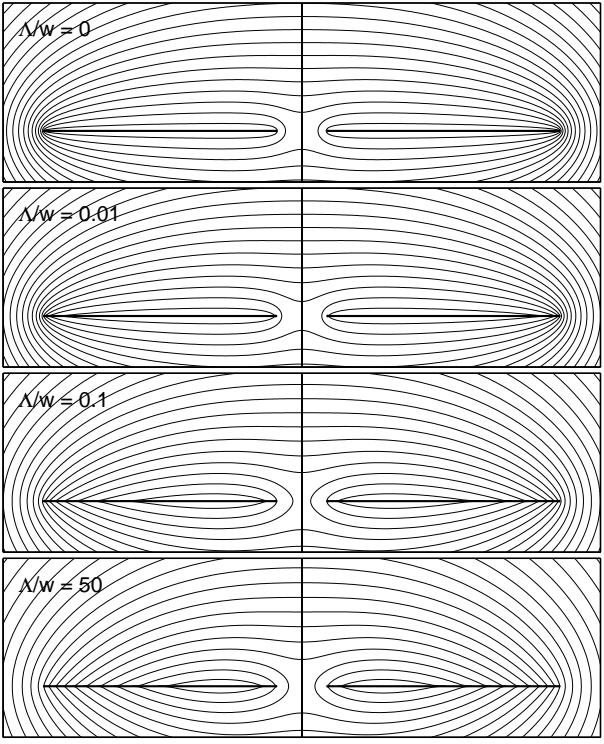}
\caption{%
  Magnetic field lines in the {\it equal-current case} for
$a/w = 0.1$ and $\Lambda/w =$ 0, 0.01, 0.1, and 50 (or $\infty$).}
\label{fig5}
\end{figure}
\begin{figure}
\includegraphics[width=8cm]{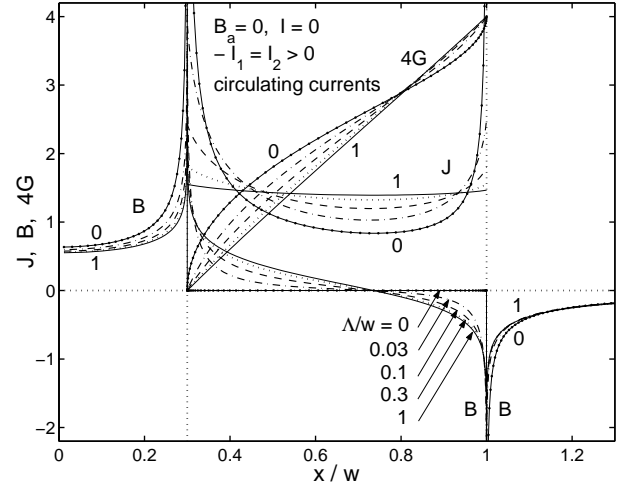}
\caption{%
  Profiles of $J_d(x)$, Eq.~(\ref{Jcircyi}), $G_d(x)$,
 Eq.~(\ref{Gcirci}), and magnetic induction $B_d(x)$ for
the {\it circulating-current case} ($B_a=0$, $-I_1=I_2>0$).
Shown are the examples $a/w=0.3$ with $\Lambda/w =$ 0 (solid lines
with dots), 0.03 (dot-dashed lines), 0.1 (dashed lines), 0.3
(dotted lines), and 1 (solid lines). $B/\mu_0$ and $J$ are
in units $I_2 /w$ and $G$ in units $I_2$.}
\label{fig6}
\end{figure}
\begin{figure} 
\includegraphics[width=8cm]{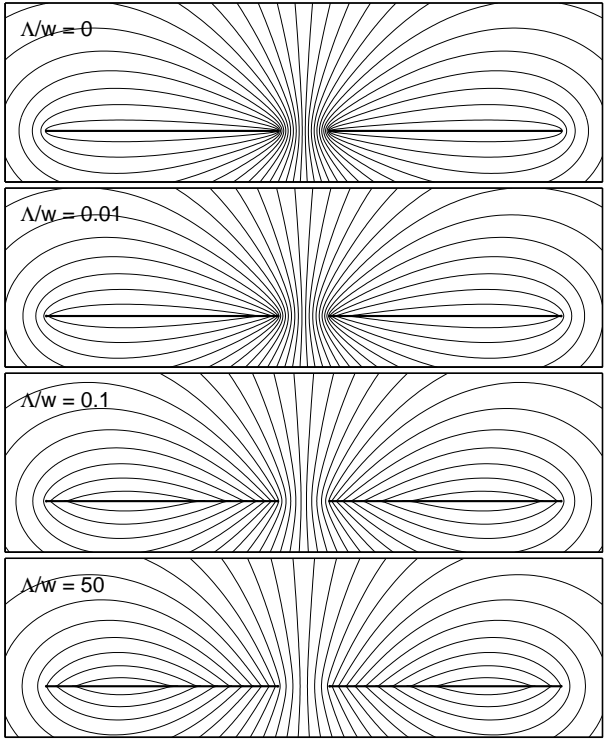}
\caption{%
  Magnetic field lines in the {\it circulating-current case} for
$a/w = 0.1$ and $\Lambda/w =$ 0, 0.01, 0.1, and 50 (or $\infty$).}
\label{fig7}
\end{figure}
\begin{figure}
\includegraphics[width=8cm]{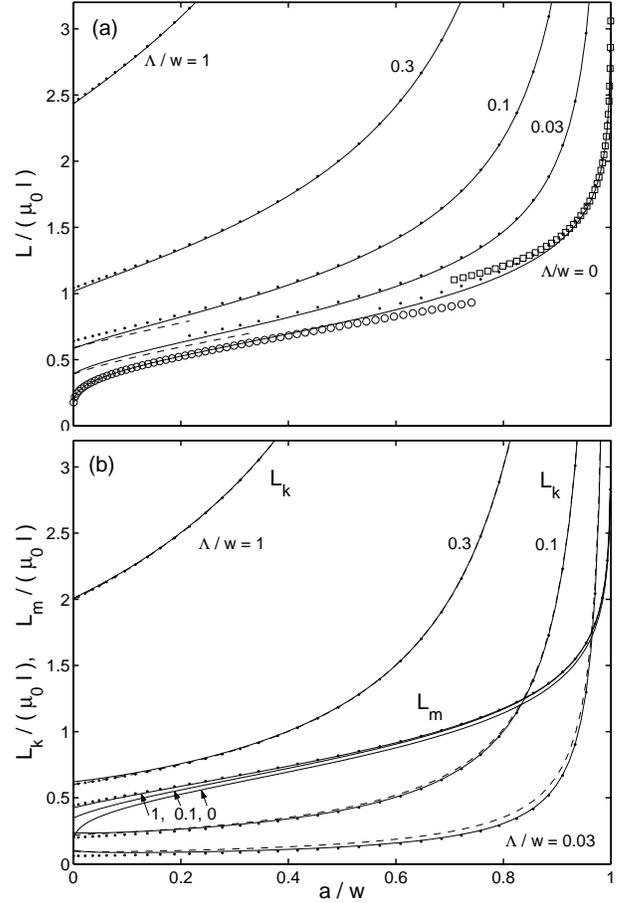}
\caption{%
(a)  Solid curves show the inductance  $L=L_{\rm m}+L_{\rm k}$
[Eq.~(\ref{Li})] vs
$a/w$ calculated for
$\Lambda/w =$ 0, 0.03, 0.1, 0.3, and 1.  See the text for descriptions of
analytic approximations shown by the open circles, open squares, dots,
and dashes. 
(b) Solid curves show the geometric inductance $L_{\rm m}$ 
[Eq.~(\ref{Lmi})] vs $a/w$ for $\Lambda/w =$ 0, 0.1, and 1 and the
kinetic inductance $L_{\rm k}$ [Eq.~(\ref{Lki})] vs $a/w$ for $\Lambda/w
=$ 0.03, 0.1, 0.3, and 1.  See the text for descriptions of
analytic approximations shown by the dotted and dashed 
curves.}
\label{fig8}
\end{figure}
\begin{equation}  
\sum_{k=1}^N K^{\rm sy}_{ik}(w_jQ^{\rm sy}_{kj}+\Lambda
\delta_{kj}) = \delta_{ij},
\label{Ksyij}
\end{equation} and applying it to Eq.\ (\ref{Ci}), we obtain
\begin{equation}  
J_I(x_i) =
\frac{I}{2\pi}\ln\Big(\frac{b}{C}\Big)
\sum_{j=1}^N K^{\rm sy}_{ij}.
\label{JIyi}
\end{equation} Since $I=2\sum_{i=1}^N w_i J_I(x_i)$, we find
\begin{equation}  
\ln\Big(\frac{b}{C}\Big)= \pi\Big/ \sum_{i=1}^N\sum_{j=1}^N w_i
K^{\rm sy}_{ij},
\label{lnbCi}
\end{equation} such that
\begin{equation}  
   J_I(x_i) = \frac{I}{2} \sum_{j=1}^N K^{\rm sy}_{ij}\Big/
\sum_{k=1}^N \sum_{l=1}^N w_k K^{\rm sy}_{kl}.
\label{JIyifinal}
\end{equation} It is remarkable that although the parameter $b$
appears in Eq.\ (\ref{Qsyij}), the final result for $J_I(x_i)$
in Eq.\ (\ref{JIyifinal}) does not depend upon
$b$. The stream function $G_I(x)$ can be evaluated as\cite{note1}
\begin{equation}  
G_I(x_i) =  \sum_{j=1}^i\sum_{k=1}^Nw_j
K^{\rm sy}_{jk}\Big/
\sum_{l=1}^N \sum_{m=1}^N w_l K^{\rm sy}_{lm}.
\label{GIi}
\end{equation}
Shown in Fig.\ 4 are plots of $J_I(x)$, $G_I(x)$, and the
corresponding magnetic induction $B_I(x)$ vs $x$ for $a/w =0.3$ and
various values of $\Lambda/w =$ 0, 0.03, 0.1, 0.3, 1.
The curves for $\Lambda=0$ exactly coincide with the analytic
expressions of Appendix A. The magnetic field lines for this case
are depicted in Fig.~5.

For the {\em circulating-current case}, Eq.\ (\ref{Fcirc/2l})
becomes
\begin{equation}  
\frac{L I_{\rm d}}{2 l} = \mu_0 \sum_{j=1}^N (w_j Q^{\rm
as}_{ij}+\Lambda
\delta_{ij})J_{\rm d}(x_j)
\label{Fcirc/2li}
\end{equation} where
   \begin{eqnarray}  
   Q^{\rm as}_{ij} &=&\frac{1}{ 2\pi} \ln\frac{x_i+x_j}
     { |x_i-x_j|},  ~~ i\ne j \,, \nonumber \\
   Q^{\rm as}_{ii} &=&\frac{1}{ 2\pi} \ln\frac{4\pi x_i}{w_i} \,,
\label{Qasij}
     \end{eqnarray} The superscript (as) is a reminder that this is
for an asymmetric current distribution [$J_{\rm d}(-x) =
-J_{\rm d}(x)$]. Defining
$K^{\rm as}_{ij} = (w_j Q^{\rm as}_{ij}+\Lambda \delta_{ij})^{-1}$,
such that
\begin{equation}  
\sum_{k=1}^N K^{\rm as}_{ik}(w_jQ^{\rm as}_{kj}+\Lambda
\delta_{kj}) = \delta_{ij},
\label{Kasij}
\end{equation} applying it to Eq.\ (\ref{Fcirc/2li}), and noting
that $I_{\rm d}=\sum_{i=1}^N w_i J_{\rm d}(x_i)$, we obtain
\begin{equation}  
J_{\rm d}(x_i) = \alpha I_{\rm d}
\sum_{j=1}^N K^{\rm as}_{ij}
\label{Jcircyi}
\end{equation} and
\begin{equation}  
L=2\alpha\mu_0 l ,
\label{Li}
\end{equation} where
\begin{equation}  
\alpha=1 \Big/ \sum_{i=1}^N\sum_{j=1}^N w_i K^{\rm as}_{ij}.
\label{alphai}
\end{equation} The stream function $G_{\rm d}(x)$ can be evaluated
as\cite{note1}
\begin{equation}  
G_{\rm d}(x_i) =
\alpha\sum_{j=1}^i\sum_{k=1}^Nw_j K^{\rm sy}_{jk}.
\label{Gcirci}
\end{equation}

Shown in Fig.\ 6 are plots of $J_{\rm d}(x)$, $G_{\rm d}(x)$,
and $B_{\rm d}(x)$ vs $x$ for $a/w =0.3$ and various values of
$\Lambda/w$. Note that these curves look similar to those in
Fig.~4, but they all have opposite parity, as can be seen from
the different profiles $B(x)$ near $x=0$. The magnetic field
lines for this case are shown in Fig.~7.

As discussed in Sec.\ II, when a vortex is present in the
region $a < |x| < w$, the sensitivity of the SQUID's critical current
$I_c$ is proportional to the magnitude of  $d\Phi_{\rm v}/dx = \phi_0
dG_{\rm d}/dx = \phi_0 J_{\rm d}(x)/I_{\rm d}$.  From Fig.\ 6 we see that
when $\Lambda \ll w$, this sensitivity is greatly enhanced when the vortex
is close to the edges $a$ and $w$ but that when $\Lambda \ge w$,  the
sensitivity is nearly independent of position.

Shown as the solid curves in Fig.\ 8(a) are plots of the inductance $L$ vs
$a/w$ for various values of $\Lambda/w =$ 0, 0.03, 0.1, 0.3, and 1.
The solid curves in Fig.\ 9(a) show the same $L$ vs $\Lambda/w$ (range
0.0045 to 2.2) for several values of $a/w =$ 0.01,
0.1, 0.4, 0.8, 0.95, and 0.99.

The geometric and kinetic contributions $L_{\rm m}$ and $L_{\rm k}$
can be calculated separately from Eqs.\ (\ref{Lm}) and
(\ref{Lk})
\begin{eqnarray}  
L_{\rm m} &=& \frac{2 \mu_0 l}{I^2_{\rm d}}
\sum_{i=1}^N\sum_{j=1}^N w_iw_jQ_{ij}^{\rm as}J_{\rm d}(x_i)J_{\rm d}(x_j)\,,
\label{Lmi}
\\ L_{\rm k} &=& \frac{2 \mu_0 l  \Lambda}{I^2_{\rm d}}
\sum_{i=1}^N\sum_{j=1}^N w_iJ^2_{{\rm d}}(x_i)\,,
\label{Lki}
\end{eqnarray} using Eqs.\ (\ref{Jcircyi}) and (\ref{alphai}).
We can show that $L_{\rm m} +L_{\rm k} = L$ using the property
of inverse matrices that
$\bm M \cdot \bm M^{-1} = \bm M^{-1} \cdot \bm M =\bm I,$ where
$\bm I$ is the identity matrix.

Shown as solid curves in Fig.\ 8(b) are $L_{\rm m}$  and
$L_{\rm k}$  vs $a/w$.
For $\Lambda=0$, when  $L_{\rm k}=0$, 
$L=L_{\rm m}$ exactly coincides with Eq.\ (A10), which may be approximated
by Eq.\ (A11) for $a/w < 0.7$ [open circles in Fig.\ 8(a)] and
by Eq.\ (A12) for $a/w > 0.7$
[open squares in Fig.\ 8(a)].
The dotted curves in Fig.\ 8(b) for $L_{\rm m}$ and $L_{\rm k}$ 
are those of Eqs.\ (C3) and (C6) in the limit $\Lambda/w \rightarrow
\infty$, when the circulating current density is uniform.
The dotted curves in Fig.\ 8(a) are obtained from $L = L_{\rm m}
+L_{\rm k}$ using the approximations of  Eqs.\ (C3) and (C6); they
are an excellent approximation to $L$ for $\Lambda/w \ge 0.03$ except for
small values of $a/w$.  
Improved agreement for small values of $\Lambda/w$ and $a/w$ 
is shown by
the dashed curves in Fig.~8(a), which show the approximation of Eq.~(B7)
for
$L$, and in Fig.~8(b), which show the approximation of Eq.~(B12) for
$L_{\rm k}$.

The solid curves in Fig.~9(b) show $L_{\rm m}$  and
$L_{\rm k}$  vs $\Lambda/w$.
The geometric inductance  $L_{\rm m}$ depends upon $\Lambda$ but only
weakly, varying slowly between its $\Lambda = 0$ asymptote [Eq.~(A10),
horizontal dot-dashed line] and its $\Lambda = \infty$ asymptote
[Eq.~(C3), horizontal dotted line].  For larger values of of $a/w$, 
$L_{\rm m}$ is nearly independent of $\Lambda$.
On the other hand, the kinetic inductance $L_{\rm m}$,  is approximately
proportional to $\Lambda/w$.
The straight dotted lines in Fig.~9(b),
calculated from the large-$\Lambda$ approximation given in Eq.~(C6),
are a good approximation to $L_{\rm m}$ except for small values of
$\Lambda/w$ and $a/w$.
The dotted curves in Fig.~9(a) are obtained from $L = L_{\rm m}
+L_{\rm k}$ using the approximations of  Eqs.\ (C3) and (C6).
Improved agreement for  small values of $\Lambda/w$ and $a/w$ is shown by
the dashed curves in Fig.~9(a), which show the approximation of Eq.~(B7)
for
$L$, and in Fig.~9(b), which show the approximation of Eq.~(B12) for
$L_{\rm k}$.

\begin{figure}
\includegraphics[width=8cm]{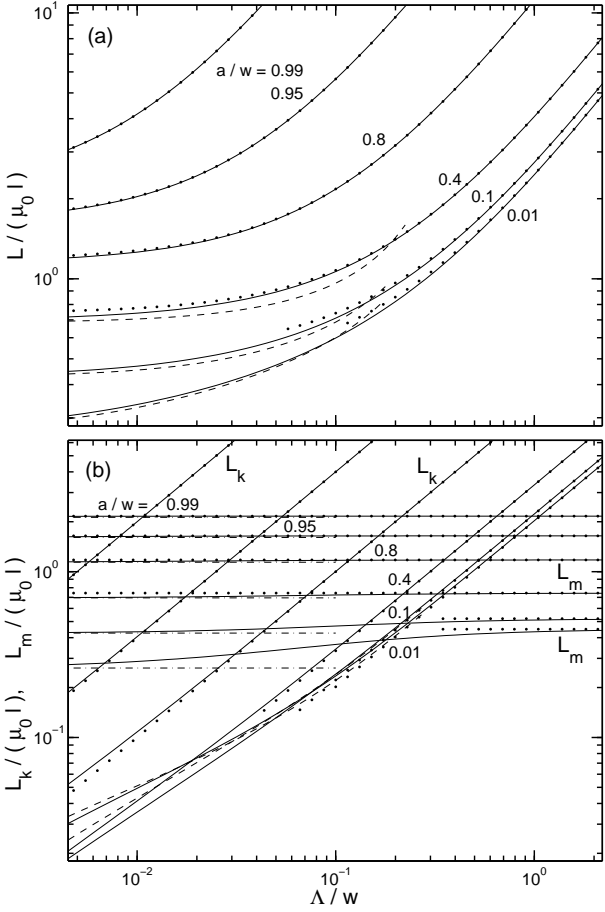}
\caption{%
(a) Solid curves show the inductance $L=L_{\rm m}+L_{\rm k}$
[Eq.~(\ref{Li})] vs $\Lambda/w$ for  $a/w =$ 0.01, 0.1, 0.4, 0.8,
0.95, and 0.99.  See the text for descriptions of analytic
approximations shown by the dotted and dashed curves.
(b) Solid curves show both the geometric inductance $L_{\rm m}$ 
[Eq.~(\ref{Lmi})] and the
kinetic inductance $L_{\rm k}$ [Eq.~(\ref{Lki})] vs $\Lambda/w$ for  $a/w
=$ 0.01, 0.1, 0.4, 0.8, 0.95, and 0.99.  See the text for descriptions of
analytic approximations shown by the dotted,  dashed,
and dot-dashed curves.}
\label{fig9}
\end{figure}
\begin{figure}
\includegraphics[width=8cm]{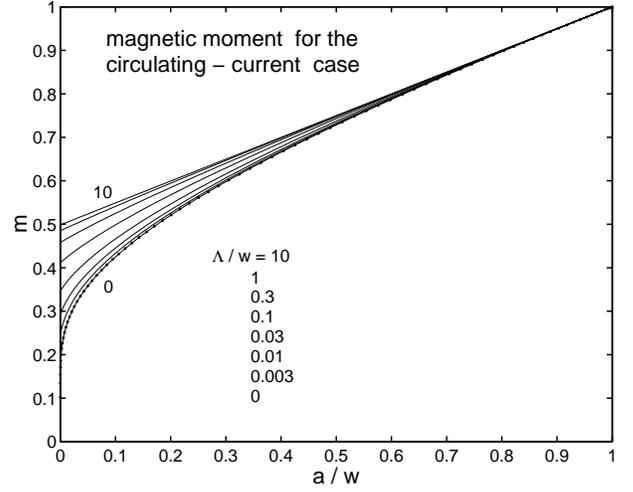}
\caption{%
  The magnetic moment $m_d$ for the {\it circulating current case}
($B_a=0$, $-I_1=I_2 =I_d >0$) plotted versus
$a/w$ for various values of $\Lambda/w$ in units $2w l I_d$.
These curves coincide with those in Fig.~13 below.}
\label{fig10}
\end{figure}
\begin{figure}
\includegraphics[width=8cm]{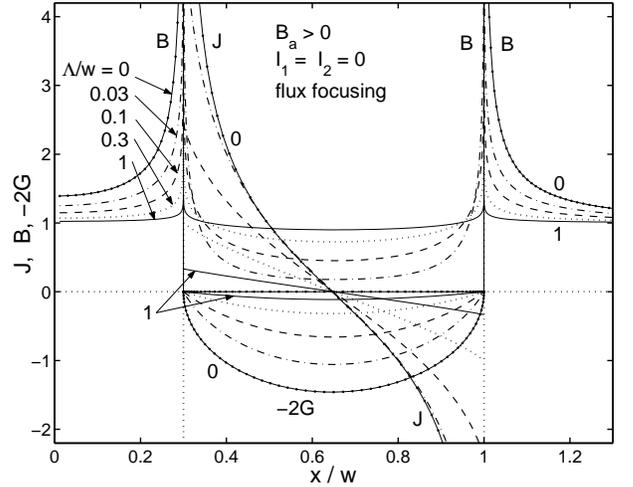}
\caption{%
  Profiles $J_f(x)$, Eq.~(\ref{Jayi}), $G_f(x)$, Eq.~(\ref{Gai}),
and magnetic induction $B_f(x)$ for the {\it flux-focusing case}
($B_a>0$, $I_1=I_2=0$).
Shown are the examples $a/w=0.3$ with $\Lambda/w =0$ (solid lines
with dots), 0.03 (dot-dashed lines), 0.1 (dashed lines), 0.3
(dotted lines), and 1 (solid lines). $B$ and $\mu_0 J$ are in
units $B_a$, and $G$ in units $w B_a/\mu_0$.}
\label{fig11}
\end{figure}
\begin{figure}
\includegraphics[width=8cm]{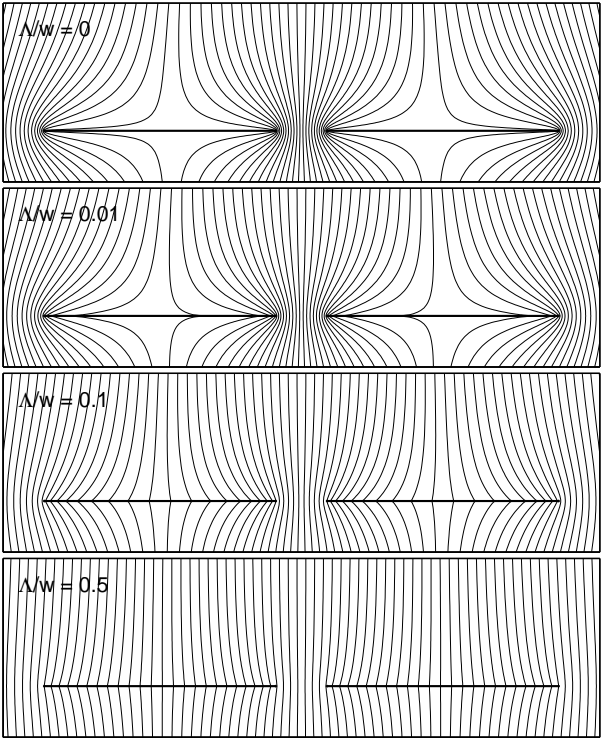}
\caption{%
  Magnetic field lines in the {\it flux-focusing case} for
 $a/w = 0.1$ and $\Lambda/w =$ 0, 0.01, 0.1, and 0.5.}
\label{fig12}
\end{figure}
\begin{figure}
\includegraphics[width=8cm]{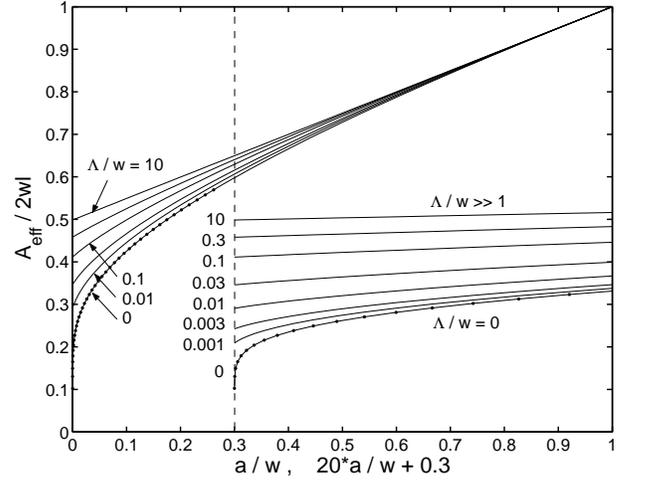}
\caption{%
  The effective area $A_{\rm eff}$, Eq.~(\ref{Aeffi}), plotted
versus the gap half width $a/w$ for several values of
$\Lambda/w =$ 0, 0.001, 0.003, 0.01, 0.03, 0.1, 0.3, and 10.
The lower-right curves show the same data shifted and stretched along
$a/w$. The dots show the exact result (A19) in the limit $\Lambda/w \to
0$. For
$\Lambda/a \ge 10$ one has $A_{\rm eff}/2wl \approx (1+a/w)/2$, Appendix
C. These curves coincide with Fig.~10 since $A_{\rm eff} = m_d/I_d$.}
\label{fig13}
\end{figure}
\begin{figure}
\includegraphics[width=8cm]{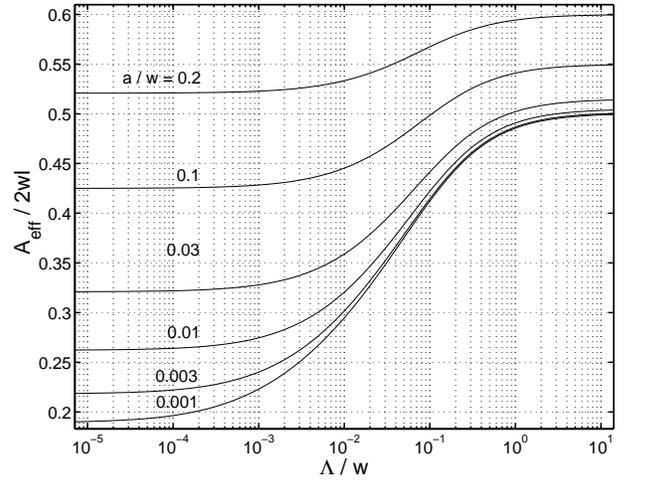}
\caption{%
  The effective area $A_{\rm eff}$, Eq.~(\ref{Aeffi}), plotted
versus $\Lambda/w$ (range $7\cdot 10^{-6}$ to 14)
for several values of $a/w =$ 0, 0.001, 0.003, 0.01, 0.03, 0.1,
and 0.2. Same data as in Fig.~13.}
\label{fig14}
\end{figure}
\begin{figure}
\includegraphics[width=8cm]{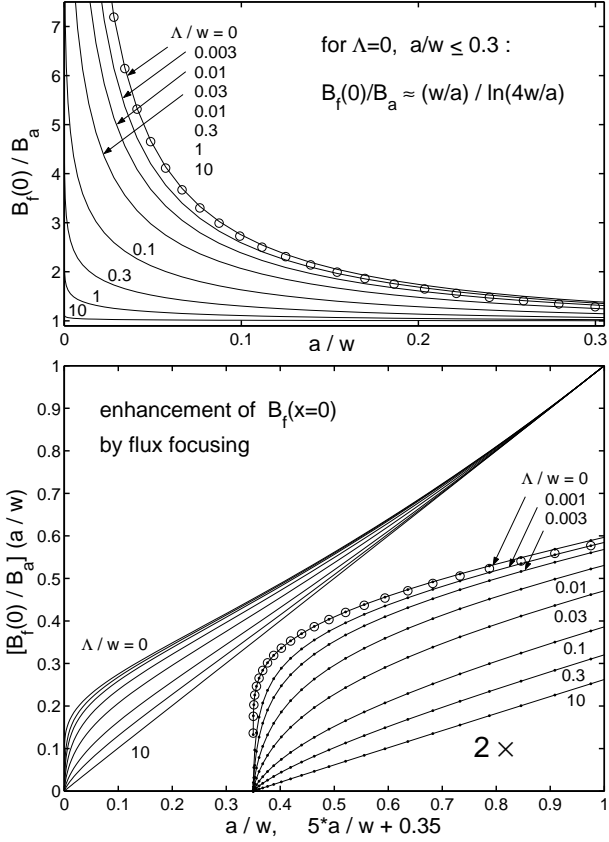}
\caption{%
  The minimum of the magnetic induction in the {\it flux-focusing
case}, $B_{\rm f}(0)=B_{\rm f}(x=0)$, referred to the applied
field $B_a$ and plotted versus the half gap width $a/w$.
Top: The ratio $B_{\rm f}(x=0)/B_a$, tending to unity for
$a/w \to 1$ and for $\Lambda/w \gg 1$, and diverging
for $a/w \to 0$ when $\Lambda = 0$.
Bottom: The same ratio multiplied by $a/w$ to avoid this
divergence and fit all data into one plot. The lower right
plot depicts the small-gap data two times enlarged along the ordinate, 
and
shifted and five times stretched along the abscissa. The circles show the
approximation $B_f(0)/B_a \approx (w/a)/\ln(4w/a)$ good for
$a/w \le 0.3$. \cite{MB1}}
\label{fig15}
\end{figure}
\begin{figure}
\includegraphics[width=8cm]{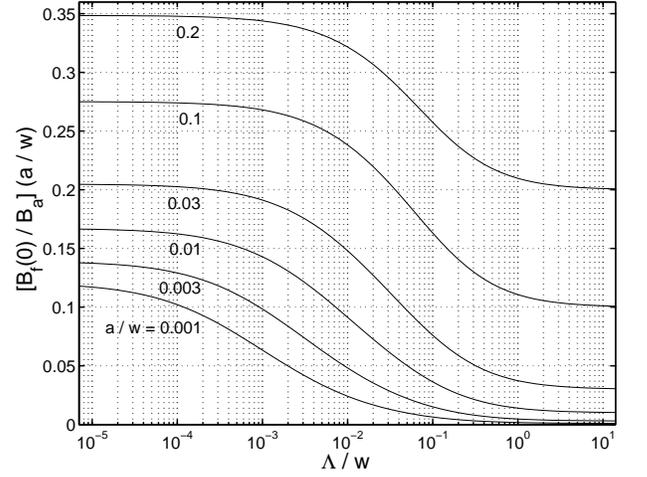}
\caption{%
  The minimum field $B_{\rm f}(0)=B_{\rm f}(x=0)$
for the {\it flux-focusing case}
as in Fig.~15 but plotted versus $\Lambda/w$  (range
$7\cdot 10^{-6}$ to 14) for several values of
$a/w =$ 0, 0.001, 0.003, 0.01, 0.03, 0.1, and 0.2.}
\label{fig16}
\end{figure}
\begin{figure}
\includegraphics[width=8cm]{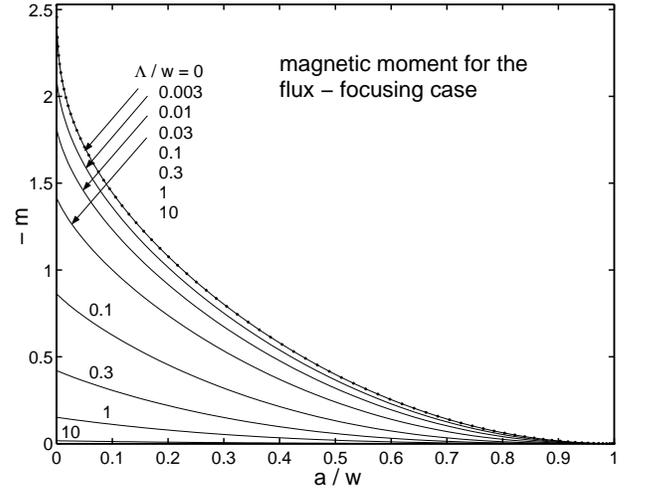}
\caption{%
  The magnetic moment $m_f$ for the {\it flux-focusing case}
plotted versus $a/w$ for various values of $\Lambda/w$
in units $w^2 l B_a/\mu_0$.}
\label{fig17}
\end{figure}
\begin{figure}
\includegraphics[width=8cm]{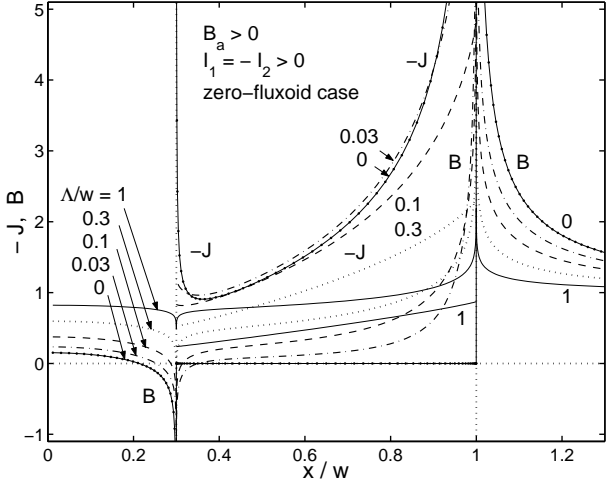}
\caption{%
  Profiles of the sheet-current density $J(x)$ [second term in
Eq.~(\ref{Jayi})]  and magnetic induction $B(x)$ generated  when a
perpendicular magnetic induction
$B_{\rm a}$ is applied in the {\it zero-fluxoid cas}e when $\Phi =
0$, 
$I_1=-I_2 >0$, and
$I=0$. Shown are the examples $a/w=0.3$ with $\Lambda/w =$ 0 (solid lines
with dots), 0.03 (dot-dashed lines), 0.1 (dashed lines), 0.3
(dotted lines), and 1 (solid lines). $B$ and $\mu_0 J$ are in
units $B_a$.}
\label{fig18}
\end{figure}
\begin{figure}
\includegraphics[width=8cm]{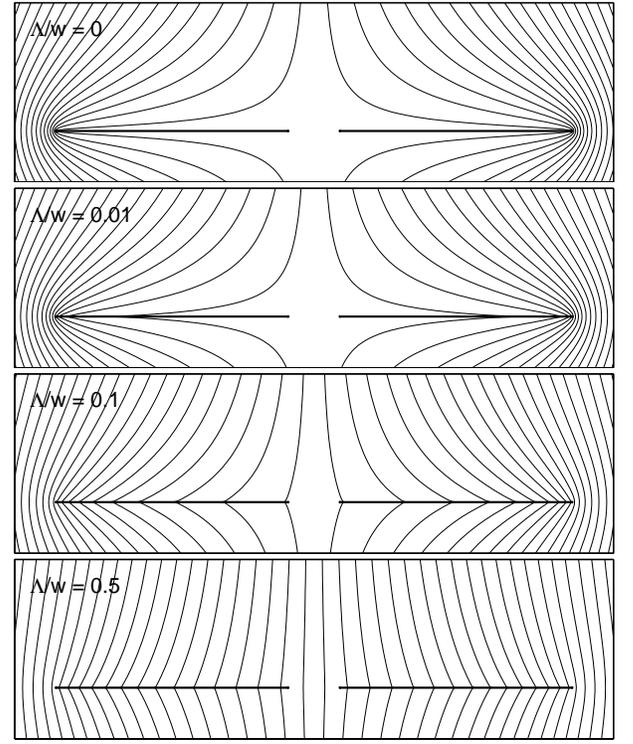}
\caption{%
  Magnetic field lines in the {\it zero-fluxoid case} $\Phi = 0$,  $B_a >0$,
and $I=0$
 for $a/w = 0.1$ and
$\Lambda/w =$ 0, 0.01, 0.1, and 0.5.}
\label{fig19}
\end{figure}
\begin{figure}
\includegraphics[width=8cm]{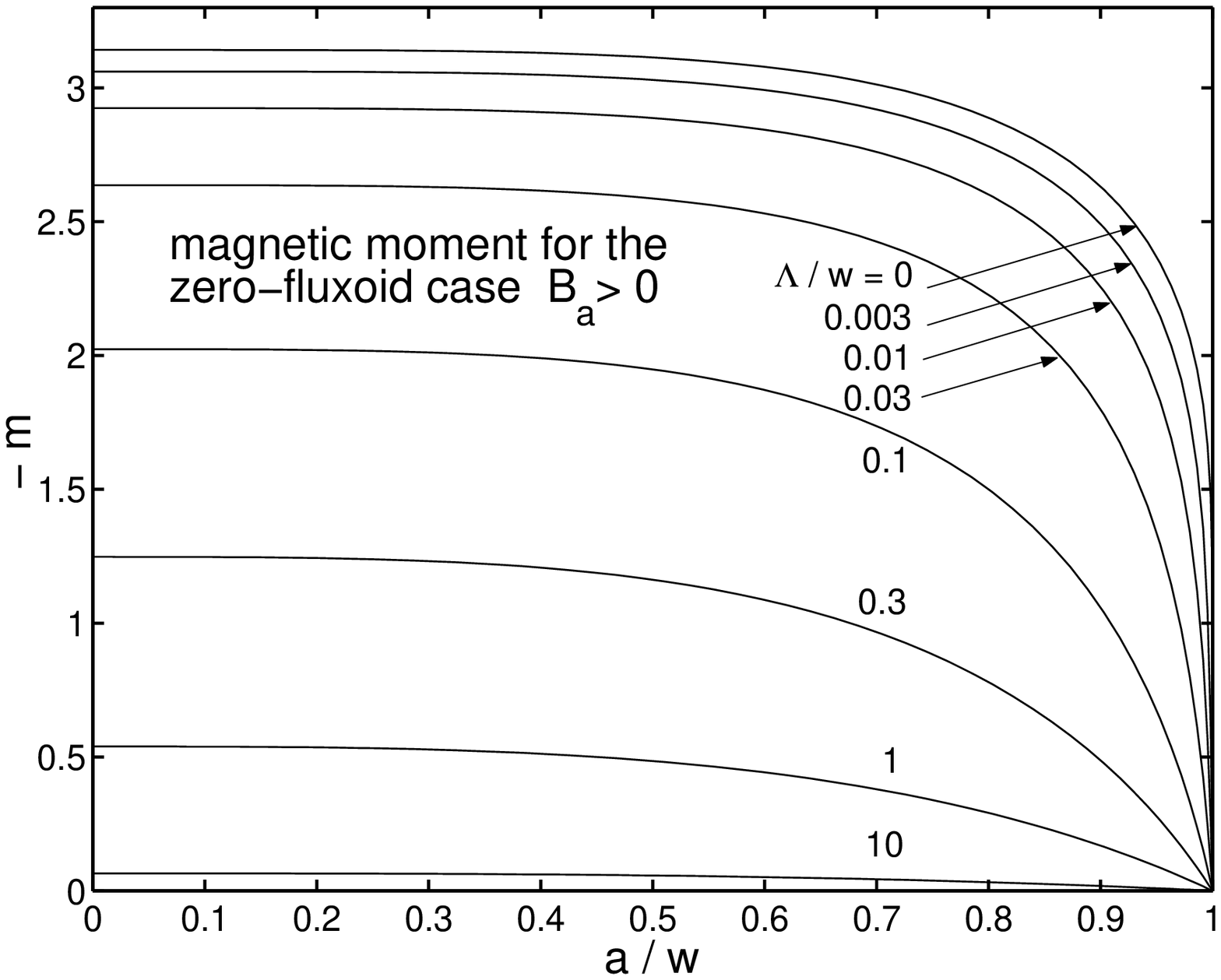}
\caption{%
  The magnetic moment $m$ for the {\it zero-fluxoid case} $\Phi = 0$,
$B_a>0$, and
$I=0$ 
plotted versus $a/w$ for various values of $\Lambda/w$
in units $w^2 l B_a/\mu_0$. At $a=\Lambda=0$ one has
$-m = \pi w^2 l B_a/\mu_0$.}
\label{fig20}
\end{figure}

In Eq.\ (3) of Ref.\ \onlinecite{Yoshida92},  Yoshida et al.\
derived an approximate expression for the kinetic inductance when
$\Lambda/w \ll 1$. We have found that their expression for $L_{\rm k}$ is
not an accurate approximation to our exact numerical results.  To
eliminate the logarithmic divergences due to the inverse square-root
dependence of the current density near the edges,  Yoshida et al.\
followed an approach used by Meservey and Tedrow,\cite{Meservey69} and
chose a cutoff length of the order of
$d$, the film thickness.  When
$d <
\lambda$, however, this approach cannot be correct, because the equations
describing the fields and currents in superconducting strips contain only
the two-dimensional screening length
$\Lambda =
\lambda^2/d$.  The cutoff length therefore
must instead be chosen to be of the order of
$\Lambda$, as we have done in Appendix B.

The magnetic moment in the $z$ direction generated by the
circulating current is, from Eq.\ (\ref{mcirc}),
\begin{equation}  
m_{\rm d} = 2\sum_{i=1}^N w_i x_i J_{\rm d}(x_i).
\label{mcirci}
\end{equation}
As shown in Fig.~10, this magnetic moment vanishes very slowly
when the gap width and $\Lambda$ go to zero, $a/w \to 0$ and
$\Lambda/w \to 0$. This can be explained by the fact that for
$\Lambda=0$ and $a < x \ll w$ one has $J_d(x) \propto 1/x$,
Eq.~(A5). The contribution of these small $x$ to $m_d$,
Eq.~(\ref{mcirci}), stays finite due to the factor $x$,
but the total current $I_d$ to which $m_d$ is normalized,
diverges when $a/w \to 0$, thus suppressing the plotted
ratio $m_d / I_d$. Interestingly, the curves in Fig.~10
coincide with those in Fig.~13; see below.
Expressions for $m_{\rm d}$ in the limits $\Lambda/w \rightarrow 0$ and 
$\Lambda/w \rightarrow \infty$ are given in Eqs.~(A13) and (C7).

For the {\em flux-focusing case},  Eq.\ (\ref{BaAeff}) becomes
\begin{equation}  
B_{\rm a } (A_{\rm eff}/2l - x_i) = \mu_0
\sum_{j=1}^N(w_j Q_{ij}^{\rm as}+\Lambda \delta_{ij})J_{\rm f}(x_j).
\label{BaAeffi}
\end{equation} Applying Eq.\ (\ref{Kasij}), we obtain
\begin{equation}  
J_{\rm f}(x_i) = \frac{B_{\rm a }}{\mu_0}
\Big( \frac{A_{\rm eff}}{2l} \sum_{j=1}^N K^{\rm as}_{ij} -
\sum_{j=1}^N K^{\rm as}_{ij}x_j\Big),
\label{Jayi}
\end{equation} where, since $\sum_{i=1}^N w_i J_{\rm f}(x_i)=0$,
the effective area is
\begin{equation}  
A_{\rm eff}=2 \alpha l
\sum_{i=1}^N\sum_{j=1}^N w_i K^{\rm as}_{ij} x_j.
\label{Aeffi}
\end{equation} The stream function $G_{\rm f}(x)$ can be evaluated
as\cite{note1}
\begin{equation} G_{\rm f}(x_i) =  \sum_{j=1}^iw_jJ_{\rm f}(x_j).
\label{Gai}       
\end{equation}

Shown in Fig.\ 11 are plots of the flux-focusing $J_{\rm f}(x)$,
$G_{\rm f}(x)$, and $B_{\rm f}(x)$ vs $x$ for $a/w =0.3$ and
$\Lambda/w =$ 0, 0.03, 0.1, 0.3, and 1.
The first term in Eq.~(\ref{Jayi}) equals the circulating-current
sheet-current density, Eq.~(\ref{Jcircyi}), with appropriate weight factor
such that the total circulating current vanishes, $I_1=I_2=0$.
The corresponding magnetic field lines are depicted in Fig.~12.
Shown in Figs.\ 13 and 14  are plots of the effective area $A_{\rm
eff}(a/w,\Lambda/w)$ versus $a/w$ and $\Lambda/w$, respectively, in units
of the maximum possible area $2wl$.  In the limit $\Lambda/w
\rightarrow 0$,  $A_{\rm
eff}$ is given by  Eq.~(A19), and when $\Lambda/w \rightarrow
\infty$, $A_{\rm
eff} = l(w+a).$ Note in Fig.\ 14 that $A_{\rm eff}$ increases
with increasing
$\Lambda$, particularly for small gap widths
$2a$. Flux focusing is reflected by the fact that for small  $a/w \to 0$
the effective area $A_{\rm eff}$ of the gap tends to a constant,
except in the limit $\Lambda \to 0$, where it vanishes very slowly,
$A_{\rm eff}/2wl \approx (\pi/2)/\ln(4w/a)$ [Eq.~(A19)]. When $a/w \to 0$,
the enhancement factor $A_{\rm eff}/2al \to \infty$ and thus  diverges
even for $\Lambda=0$. In the limit $a/w \ll 1$,
$A_{\rm eff}(0, \Lambda/w)$ tends to a universal function
[see Fig.~14]. Interestingly, Figs.~13 and 10 show identical curves; this
is because the identity $A_{\rm eff} =m_d / I_d$ holds
for all values of $a/w$ and $\Lambda/w$, as proved in general
in Sec.~II.

  Figures 15 and 16 show the minimum of the magnetic induction
in the {\it flux-focusing case}, $B_{\rm f}(0)=B_{\rm f}(x=0)$ [see
Fig.~11], plotted versus $a/w$ and $\Lambda/w$, respectively.
The ratio $B_{\rm f}(0) / B_a \ge 1$  tends to unity for
$a/w \to 1$ and for $\Lambda/w \gg 1$, and it diverges
for $a/w \to 0$ when $\Lambda = 0$. The curve for $\Lambda=0$
exactly coincides with the analytic expression
$B_f(0)/B_a = w {\bm E}(k')/ a {\bm K}(k')$ obtained from Eq.~(A17). For
$a/w \ll 1$ this yields $B_f(0)/B_a \approx (w/a)/\ln(4w/a)$,
which is a good approximation for $0 < a/w \le 0.3$. \cite{MB1}
The  magnetic moment $m_{\rm f}$ for the flux-focusing case,
calculated from Eq.\ (\ref{mcirci}) but with $J_{\rm d}(x_i)$
replaced by $J_{\rm f}(x_i)$,  is shown in Fig.~17.  Expressions for 
$m_{\rm f}$ in the limits $\Lambda/w \to 0$ and $\Lambda/w \to \infty$
are given in Eqs.~(A20) and (C9)

Shown in Fig.\ 18 are profiles for the {\it zero-fluxoid case} with plots
of
$J(x)$ and the corresponding $B(x)$
generated by an applied magnetic induction $B_a>0$ when the junctions are
short-circuited such that $\Phi = 0$ and  $I_1=-I_2 >0$; for comparison see
analogous profiles in Sec. 2.5 of Ref.\
\onlinecite{Babaei02} for two parallel strips and in Sec.\ IV of
Ref.\
\onlinecite{Babaei03} and Sec.~4 of Ref.\ \onlinecite{Brandt04} for rings.
That the current density $J(x)$ in the zero-fluxoid case is given by the
second term on the right-hand sides of Eqs.\ (\ref{Jay}) and
(\ref{Jayi}), can be seen by setting $\Phi_{\rm f} =
B_{\rm a} A_{\rm eff} = 0$ in Eqs. (\ref{BaAeff}), (\ref{Jay}),
(\ref{BaAeffi}), and (\ref{Jayi}). 
 Depicted  in Fig.\ 18 are the examples
$a/w=0.3$ with $\Lambda/w =$ 0, 0.03, 0.1, 0.3, and 1.
Figure 19 shows the magnetic field lines for this case and
Fig.~20 the magnetic moment $m$.  Expressions for $J$, $B$, and $m$ for
the zero-fluxoid case in the limits  $\Lambda/w \to 0$ and 
$\Lambda/w \to \infty$
are given in Appendixes A and C.

\section{Summary}

In Sec.\ II of this paper we have presented general equations governing
the static behavior of a thin-film dc SQUID for all  values of the
Pearl length $\Lambda = \lambda^2/d$, where the London
penetration depth $\lambda$ is larger than $d$, the film thickness.  The
SQUID's critical current
$I_c$ depends upon the effective flux
$\Phi$, which is the sum of the magnetic flux up through a contour
surrounding the central hole and a term proportional to the line integral
of the current density around this contour.  For a
symmetric SQUID there are three important contributions to $\Phi$:  
a circulating-current term $\Phi_{\rm d}$,
a vortex-field term $\Phi_{\rm v}$,
and a flux-focusing term
$\Phi_{\rm f}$, all of which depend upon $\Lambda$.  Since $\Lambda$ is a
function of temperature, an important consequence is that all of the
contributions to $\Phi$ are temperature-dependent.

The circulating-current term $\Phi_{\rm d}$ can be expressed in terms of 
the SQUID inductance $L$ and the circulating current
$I_{\rm d}$ via 
$\Phi_{\rm d} =L I_{\rm d}$.  The SQUID inductance has two contributions, $L
= L_{\rm m} + L_{\rm k}$, where the first term is the geometric
inductance (associated with the energy stored in the magnetic field) and
the second is the kinetic inductance (associated with the kinetic
energy of the circulating supercurrent). Both contributions are functions
of
$\Lambda$, since they both depend on the spatial distribution of the
current density. However,
$L_{\rm m}$ depends only weakly upon
$\Lambda$, because for the same circulating current $I_{\rm d}$, the
energy stored in the magnetic field does not vary greatly as $\Lambda$
ranges from zero to infinity.  On the other hand, because the kinetic
energy density is proportional to
$\Lambda$, $L_{\rm k}$ is also nearly proportional to
$\Lambda$, with deviations from linearity occurring only for small
values of $\Lambda/w$.   

The vortex-field term can be written as $\Phi_{\rm v}
= \phi_0 G_{\rm d}$, where  $G_{\rm d}$ is a dimensionless stream
function describing the circulating sheet-current density ${\bm J}_d$. 
  Roughly speaking, when
$\Lambda$ is small,
$I_c$ is most strongly dependent upon the vortex position when the vortex
is close to the edges of the film, but when $\Lambda$ is large, $I_c$
is equally  sensitive to the vortex position wherever the vortex
is.  
Recent experiments\cite{Doenitz05} have used the
relationship
$\Phi_{\rm v} = \phi_0 G_{\rm d}$ 
to determine the vortex-free sheet-current density ${\bm J}_d(x,y)$ from
vortex images obtained via low-temperature scanning electron
microscopy.\cite{Straub01,Doenitz04} 
The experimental data obtained in magnetic fields up to 40 $\mu$T are in
excellent agreement with numerical calculations of ${\bm J}_d(x,y)$, 
confirming the validity of the above relationship, even in the
presence of many (up to 200) vortices in the SQUID washer.

The flux-focusing term can be expressed as 
$\Phi_{\rm f}= B_{\rm a} A_{\rm eff}$, where  $B_{\rm a}$ is the applied
magnetic induction and $A_{\rm eff}$ is the effective area of the
central hole of the SQUID.  Although $A_{\rm eff}$ is primarily
determined by the dimensions of the SQUID, it also depends upon the value
of 
$\Lambda.$

To illustrate the $\Lambda$ dependence of the above quantities, in Sec.\
III of this paper we  analyzed in detail the behavior of a long SQUID
whose central region resembles a coplanar stripline. We  numerically
calculated the profiles of the sheet-current density, stream function, and
magnetic induction in the equal-current, circulating-current,
flux-focusing, and zero-fluxoid cases for various representative
values of
$\Lambda$.  We  presented plots of the inductances $L,$ 
$L_{\rm m}$, and $L_{\rm k}$, the effective area $A_{\rm eff}$, and the
magnetic moments for these cases.  Useful analytic approximations are
provided for the
$\Lambda/w \to 0$ limit in Appendix A, for small $\Lambda/w$ and $a/w$ in
Appendix B, and for the $\Lambda/w \to \infty$
limit in Appendix C.

We are in the process of applying the above theory to square
and circular SQUIDs, using the numerical method of Ref.\
\onlinecite{Brandt05}.

\acknowledgments
We thank D. Koelle for stimulating discussions.
This work was
supported  in part by Iowa State University of
Science and Technology under Contract No.\ W-7405-ENG-82 with
the U.S.\ Department of Energy and in part by the German Israeli
Research Grant Agreement (GIF) No G-705-50.14/01.

\appendix

\section{the limit $\Lambda/w = 0$}   

In the {\em ideal-screening limit} $\Lambda/w = 0$, the $y$ component of
the sheet-current density in the strips ($a < |x| <w$) for the {\it
equal-current case} is\cite{Babaei02}
\begin{equation}   
J_I(x)  = \frac{I}{\pi} \frac{|x|}
{[(x^2-a^2)(w^2-x^2)]^{1/2}},
\label{JIyzero}
\end{equation} and the $z$ component of the magnetic induction in the
plane
$z = 0$ of the strips is
\begin{eqnarray}  
B_I(x)\! &=& \! -\frac{\mu_0 I}{2 \pi}
\frac{x}{[(x^2-a^2)(x^2-w^2)]^{1/2}}, ~|x|>w,~~~~
\label{BIzx>w} \\
      &=& 0, ~a<|x|<w,
\label{BIza<x<w}\\
      &=& \frac{\mu_0 I}{2 \pi}\,
\frac{x}{[(a^2-x^2)(w^2-x^2)]^{1/2}}, ~|x|<a,
\label{BIzx<a}
\end{eqnarray} and the constant $C$ in Eqs.\ (\ref{AJy}), (\ref{C}),
(\ref{JIyC}), and (\ref{lnbC}) is $C = \sqrt{w^2-a^2}/2.$

For the {\it circulating-current case}, the $y$ component of the
sheet-current density  in the strips ($a < |x| <w$) in the
   limit  $\Lambda/w = 0$ is\cite{Babaei02}
\begin{equation}  
J_{\rm d}(x)  = \frac{2 B_0}{\mu_0}
\frac{x}{|x|}\frac{w^2} {[(x^2-a^2)(w^2-x^2)]^{1/2}},
\label{Jcircyzero}
\end{equation} and the $z$ component of the  magnetic induction
 in the plane
$z = 0$ of the strips is
\begin{eqnarray}  
B_{\rm d}(x)\! & = &\! -B_0
\frac{w^2}{[(x^2-a^2)(x^2-w^2)]^{1/2}}, ~|x|>w,~~~~ \\
\label{Bcirczx>w}
      &=& 0, ~a<|x|<w,
\label{Bcircza<x<w}\\
      &=& B_0 \frac{w^2}{[(a^2-x^2)(w^2-x^2)]^{1/2}}, ~|x|<a,~
\label{Bcirczx<a}
\end{eqnarray} where the parameter $B_0$, the magnetic flux
$\Phi_{{\rm d}}$  in the $z$
direction in the slot, the circulating current $I_{\rm d}$,
and the geometric inductance $L_{\rm m}$ are related by
\begin{equation}  
\Phi_{{\rm d}} = L_{\rm m}I_{\rm d} = 2B_0lw\bm K(k)
\end{equation} and
\begin{equation}  
   L_{\rm m} = \mu_0l\bm K(k)/\bm K(k'),
\label{LA}
\end{equation} where $\bm K(k)$ is the complete elliptic integral
of the first kind of modulus $k = a/w$ and complementary modulus
$k' = \sqrt{1-k^2}$.
The geometric inductance is well approximated for small $a/w$ by 
\begin{equation} 
 L_{\rm m} = (\pi\mu_0l/2)/\ln(4w/a),
\label{LAsmalla}
\end{equation}
neglecting corrections proportional to $a^2/w^2$,
and for small $(w-a)/w$ by 
\begin{equation} 
 L_{\rm m} = (\mu_0l/\pi)\ln[16/(1-a^2/w^2)],
\label{LAlargea}
\end{equation}
neglecting corrections proportional to $1-a^2/w^2$.
In the limit that $\Lambda = 0$, the kinetic inductance vanishes ($L_{\rm
k} = 0)$, and the inductance in Eq.\ (\ref{LA}) becomes the
total  inductance:
$L = L_{\rm m}$.
The magnetic moment [see Eq.\ (\ref{mcirc})] can be obtained from Eqs.\
(13)-(16) of Ref.\ \onlinecite{Babaei02}:
\begin{equation} 
m_{\rm d} = [ \pi lw/{\bm K}(k')]I_{\rm d}.
\end{equation}

For the {\it flux-focusing case}, the $y$ component of the sheet-current
density  in the strips ($a < |x| <w$) in the  limit  $\Lambda/w = 0$
is\cite{Babaei02}
\begin{equation}  
J_{\rm f}(x)  = \frac{2 B_{\rm a }}{\mu_0
\bm K(k')} \frac{x}{|x|}
\frac{\bm E(k')w^2-2\bm K(k')x^2} {[(x^2-a^2)(w^2-x^2)]^{1/2}},
\label{Jayzero}
\end{equation} and the $z$ component of the magnetic induction in the
plane
$z = 0$ of the strips is
\begin{eqnarray}  
B_{\rm f}(x)\! & = &\!
-\frac{B_{\rm a }}{\bm K(k')} \frac{\bm E(k')w^2-2
  \bm K(k')x^2}{[(x^2-a^2)(x^2-w^2)]^{1/2}}, \nonumber \\ && ~|x|>w,
\label{Bazx>w}
\\
  &=& 0, ~a<|x|<w,~
\label{Baza<x<w}\\
  &=& \frac{B_{\rm a }}{\bm K(k')} \frac{\bm E(k')w^2-2\bm K(k')x^2}
    {[(a^2-x^2)(w^2-x^2)]^{1/2}}, \nonumber \\ && ~|x|<a,~
\label{Bazx<a}
\end{eqnarray} where $\bm E(k')$ is the complete elliptic integral
of the second kind of complementary modulus $k' = \sqrt{1-k^2}$ and
modulus $k = a/w$. The magnetic flux in the $z$ direction in the slot is
\begin{equation}  
\Phi_{{\rm f}} = \pi B_{\rm a } lw / \bm K(k'),
\end{equation} and the effective area $A_{\rm eff}$ of the slot is
\begin{equation} 
A_{\rm eff} = \Phi_{{\rm f}}/B_{\rm a } = \pi l
w/\bm K(k').
\end{equation}
Note that  $A_{\rm eff} = m_{\rm d}/I_{\rm d}.$
The magnetic moment generated by $J_{\rm f}(x)$ is
\begin{equation} 
m_{\rm f} = -\pi l [w^2 + a^2 - 2w^2{\bm E}(k')/{\bm K}(k')]B_{\rm
a}/\mu_0.
\end{equation}

For the {\it zero-fluxoid case}, the $y$ component of the sheet-current
density in the strips can be obtained from Sec.\ 2.5 of Ref.\
\onlinecite{Babaei02}:
\begin{equation} 
J(x)  = -\frac{2 B_{\rm a }}{\mu_0}
\frac{x}{|x|}\frac{x^2\!-\![1\!-\!{\bm E}(k)/{\bm K}(k)]w^2}
{[(x^2-a^2)(w^2-x^2)]^{1/2}}.
\label{Jzerofluxoid}
\end{equation}
The corresponding $z$ component of the magnetic induction
is\cite{Babaei02}
\begin{eqnarray} 
B(x)\! & = &\!
B_{\rm a } ~\frac{x^2\!-\![1\!-\!{\bm E}(k)/{\bm
K}(k)]w^2}{[(x^2-a^2)(x^2-w^2)]^{1/2}}, |x|>w,~~
\label{B0zx>w}
\\
  &=& 0, ~a<|x|<w,~
\label{B0za<x<w}\\
  &=& B_{\rm a}~ \frac{[1\!-\!{\bm E}(k)/{\bm
K}(k)]w^2\!-\!x^2}
    {[(a^2-x^2)(w^2-x^2)]^{1/2}}, |x|<a.~~
\label{B0zx<a}
\end{eqnarray} 
The magnetic moment generated by $J(x)$ is
\begin{equation} 
m = -\pi l [2w^2{\bm E}(k)/{\bm K}(k)- w^2 + a^2  ]B_{\rm
a}/\mu_0.
\end{equation}

\section{Behavior for small $\Lambda$ and small $a$}

In this section we present some expressions for $L$, $L_{\rm k}$, and
$L_{\rm m}$ that follow from approximating the circulating-current
distribution for small values of
$\Lambda$ and $a$.

When the slot is very narrow ($a/w \ll
1$), we approximate the sheet-current density in the region $a < x <
w$ generated by the fluxoid
$\Phi_{\rm d}$  via
\begin{equation} 
J_{\rm d}(x) =I_0
\frac{w}{\sqrt{(x^2-a^2+\delta^2)(w^2-x^2+\delta^2)}},
\label{Jdydelta}
\end{equation} 
where $I_0 = 2 \Phi_{\rm d}/\pi \mu_0 l$ and $\delta$ is a quantity of order 
$\Lambda =
\lambda^2/d$ determined as follows. When $l \rightarrow \infty$ and then
$a
\rightarrow 0$ and
$w \rightarrow
\infty$, an exact calculation yields for $x > 0$ 
\begin{equation} 
J_{y}(x) = \frac{2 \Phi_{\rm d}}{\pi \mu_0 l \Lambda}
\int_0^\infty \frac{e^{-xt/\Lambda} dt}{t^2 + 1}.
\label{Jdyzeroa}
\end{equation}
We find $\int_0^b J_{y}(x) dx =  
(2 \Phi_{\rm d}/\pi \mu_0 l)\ln(\gamma b/2 \Lambda)$ when $b \gg \Lambda$,
where $\gamma = e^C = 1.781...$, and $C = 0.577... $ is Euler's constant. 
From Eq.\ (\ref{Jdydelta}) we find 
$\int_0^b J_{\rm d}(x) dx =  
(2 \Phi_{\rm d}/\pi \mu_0 l)\ln(2 b/\delta)$ when $a = 0$ and 
$\delta \ll b \ll w$.  Comparing these two integrals we obtain $\delta =
(4/\gamma) \Lambda = 2.246  \Lambda$.  Integrating Eq.\
(\ref{Jdydelta}) from $a$ to $w$ to obtain $I_{\rm d}$, we find
\begin{equation} 
I_{\rm d} = I_0
\frac{w}{\sqrt{w^2+\delta^2}}[F(\lambda_a,q)-F(\lambda_w,q)],
\label{B3}
\end{equation}
where $F(\phi,k)$ is the elliptic integral of the first kind and 
\begin{eqnarray} 
\lambda_a &=& \arcsin{\sqrt{\frac{w^2-a^2 +\delta^2}{w^2-a^2
+2\delta^2}}}\,,\\
\lambda_w &=& \arcsin{\frac{\delta}{\sqrt{w^2-a^2 +2\delta^2}}}\,,\\
q&=&\sqrt{\frac{w^2-a^2+2\delta^2}{w^2+\delta^2}}\,.
\end{eqnarray}

Expanding Eq.\ (\ref{B3}) for $a \ll w$ and $\delta \ll w$, using $I_0 =
2 \Phi_{\rm d}/\pi \mu_0 l$, and neglecting terms of order $a^2/w^2$ and
$\delta^2/w^2$, we obtain
\begin{equation} 
L=\Phi_{\rm d}/I_{\rm d} = (\pi\mu_0
l/2)/\{\ln[4w/(a+\delta)]-\delta/w\},
\end{equation}
where $\delta = 2.246 \Lambda$.  Note that Eq.\ (B7) reduces to Eq.\
(A11) when $\Lambda = 0$.

From Eq.\ (\ref{Jdydelta}) we obtain the approximation
\begin{equation} 
\int_a^wJ_{\rm d}^2(x)dx = I_0^2\frac{w}{(w^2-a^2+\delta^2)}(f_a+f_w),
\label{intJ2}
\end{equation}
where 
\begin{eqnarray} 
f_a&=&\frac{w}{\sqrt{\delta^2-a^2}}
\tan^{-1}\frac{(w-a)\sqrt{\delta^2-a^2}}{a(w-a)+\delta^2},
\nonumber \\
&&a<\delta, \\
&=&\frac{w}{\sqrt{a^2-\delta^2}}
\tanh^{-1}\frac{(w-a)\sqrt{a^2-\delta^2}}{a(w-a)+\delta^2},
\nonumber \\
&&a>\delta, \\
f_w&=&\frac{w}{\sqrt{w^2+\delta^2}}
\tanh^{-1}\frac{(w-a)\sqrt{w^2+\delta^2}}{w(w-a)+\delta^2}.
\end{eqnarray}

Using Eqs.\ (\ref{B3}) and  (\ref{intJ2}), we
obtain from Eq.\ (\ref{Lk})
\begin{equation} 
L_{\rm k} = 2 \mu_0 l \frac{\Lambda}{w} 
\frac{(w^2+\delta^2)}{(w^2-a^2+2\delta^2)}
\frac{(f_a+f_w)}{[F(\lambda_a,q)-F(\lambda_w,q)]^2},
\label{Lkapprox}
\end{equation}
where $\delta = 2.246  \Lambda$.
Although our intention in using the ansatz of Eq.\ (\ref{Jdydelta})
initially was to obtain an improved approximation to $L_{\rm k}$ for 
small values of  $\Lambda$ and  $a$, we see  from Figs.\ 8(b) and 9(b)
that Eq.\ (\ref{Lkapprox}) provides a reasonably good approximation  for
all values of
$\Lambda$ and $a$. 

\section{the limit $\Lambda/w \rightarrow \infty$}  

In the {\em weak-screening limit} $\Lambda/w \rightarrow \infty$,
$K^{\rm sy}(x,x')=\Lambda^{-1}\delta(x-x')$, the $y$ component of
the sheet-current density in the strips ($a < |x| <w$) in the {\it
equal-current case} is uniform,
$J_I = I/2(w-a)$,  the $z$ component of the magnetic induction in the
plane of the strips obtained from the Biot-Savart law is
\begin{equation}  
B_I(x) = \frac{\mu_0 I}{4 \pi (w-a)} \ln
\Big|\frac{(x-w)(x+a)}{(x+w)(x-a)}\Big|,
\label{BIzinfty}
\end{equation} and the constant $C$ in Eqs.\ (\ref{AJy}), (\ref{C}),
(\ref{JIyC}), and (\ref{lnbC}) is $C = w\exp[-\pi\Lambda/(w-a)].$

For the {\it circulating-current case} in the limit
$\Lambda/w \rightarrow \infty$,
$K^{\rm as}(x,x')=\Lambda^{-1}\delta(x-x'),$ $\alpha =
\Lambda/(w-a)$, the $y$ component of the sheet-current density in the
strips is again uniform,
$J_{\rm d} = I_{\rm d}/(w-a)$ for $a < x < w$, and the $z$ component of
the magnetic induction
in the plane of the strips obtained from the Biot-Savart law is
\begin{equation}  
B_{\rm d}(x) = \frac{\mu_0 I_{\rm d}}{2
\pi (w-a)} \ln
\Big|\frac{x^2-w^2}{x^2-a^2}\Big|.
\label{Bizinfty}
\end{equation}
The geometric inductance is, from Eq.\ (\ref{Lm})
\begin{eqnarray} 
L_{\rm m} &=& \frac{\mu_0 l}{\pi
(w-a)^2}\Big[w^2\ln\Big(\frac{4w^2}{w^2-a^2}\Big)\nonumber \\
&&-2aw\ln\Big(\frac{w+a}{w-a}\Big)
+a^2\ln\Big(\frac{4a^2}{w^2-a^2}\Big)\Big],
\end{eqnarray}
which is independent of $\Lambda$.  Equation (C3) is well approximated for
small
$a/w$ by 
\begin{equation} 
 L_{\rm m} = (\mu_0l/\pi)(1+2a/w)\ln4,
\label{LCsmalla}
\end{equation}
neglecting corrections proportional to $a^2/w^2$,
and for small $(w-a)/w$ by 
\begin{equation} 
 L_{\rm m} = (\mu_0l/\pi)\Big[\ln\frac{2}{1-a/w}+
\frac{3}{2} -\frac{1}{2}(1-a/w)\Big],
\label{LClargea}
\end{equation}
neglecting corrections proportional to $(1-a/w)^2$.
From Eq.\ (\ref{Lk}) we obtain the  kinetic inductance 
\begin{equation} 
 L_{\rm k} = 2 \mu_0 l \Lambda/(w-a).
\label{LkC}
\end{equation}
When $\Lambda \gg w$, the total inductance $L$ is dominated
by the  kinetic inductance ($L_{\rm k} \gg  L_{\rm m}$), such that
$L \approx L_{\rm k}$.
Since $J_{\rm d}$ is uniform, the magnetic moment is easily found from
Eq.\ (\ref{mcirc}) to be
\begin{equation} 
m_{\rm d} = l(w+a)I_{\rm d}.
\end{equation}

For the {\it flux-focusing case} in the limit
$\Lambda/w \rightarrow \infty$,
$K^{\rm as}(x,x')=\Lambda^{-1}\delta(x-x'),$ the $y$ component of
the sheet-current density is
$J_{\rm f}(x) = B_{\rm a }(w+a-2x)/2\mu_0\Lambda$ for $a < x < w$,
the effective area is $A_{\rm eff} = l(w+a) = m_{\rm d}/I_{\rm d}$, and
the
$z$ component of the magnetic induction in the plane of the strips is
$B_{\rm f}(x) = B_{\rm a } +B_{s\rm f}(x)$, where from the
Biot-Savart law
\begin{eqnarray}  
B_{s\rm f}(x)& =& \frac{B_{\rm a }}{4 \pi
\Lambda} \Big[(w+a) \ln
\Big|\frac{x^2-w^2}{x^2-a^2}\Big| \nonumber\\ &+& 2x
\ln\Big|\frac{(x+w)(x-a)}{(x-w)(x+a)}\Big| -4(w-a)\Big].~
\label{bazinfty}
\end{eqnarray}
The magnetic moment generated by $J_{\rm f}(x)$ in this limit is 
\begin{equation} 
m_{\rm f} = -[l(w-a)^3/6\Lambda](B_{\rm a}/\mu_0).
\end{equation}

For the {\it zero-fluxoid case} in the limit
$\Lambda/w \rightarrow \infty$, the applied field is only weakly screened,
and the $z$ component of the magnetic flux density is nearly equal to the
applied magnetic induction, $B(x) \approx B_{\rm a}$.  The $y$ component
of the vector potential is approximately given by $A(x) = B_{\rm a}x$,
and the
$y$ component of the induced sheet-current density, obtained from Eq.\
(\ref{j}) with
$\gamma = 0$, is $J(x) = - (B_{\rm a}/\mu_0
\Lambda)x$.  To the next order of approximation, $B(x) = B_{\rm a} +
B_s(x),$ where the self-field $B_s$ is found from the Biot-Savart law
\begin{equation}  
B_s(x) = \frac{B_{\rm a }}{2 \pi
\Lambda} \Big[x
\ln\Big|\frac{(x+w)(x-a)}{(x-w)(x+a)}\Big| -2(w-a)\Big].~
\label{bzinfty}
\end{equation}
The magnetic moment generated by $J(x)$ in this limit is 
\begin{equation} 
m = -[2l(w^3-a^3)/3\Lambda] (B_{\rm a}/\mu_0).
\end{equation}


\begin{thebibliography}{99}
\bibitem{foot3} A similar relation between the vortex-generated flux and
the stream function was derived by Humphreys in Ref.\
\onlinecite{Humphreys99} using inductance arguments.
\bibitem{note1} The accuracy of this sum (error $\propto 1/N$) is
improved (error $\propto 1/N^2$) by multiplying the last term $j=i$
with $1/2$. Even higher accuracy can be reached by interpolation,
using different weights for each $i$ that vanish at the boundaries
($a,\, x_i$)  of the indefinite integral.
\end{thebibliography}
\end{document}